\documentclass[lettersize,journal]{IEEEtran}
\usepackage{amsmath,amsfonts}
\usepackage{array}
\usepackage{textcomp}
\usepackage{stfloats}
\usepackage{url}
\usepackage{subfigure}
\usepackage{verbatim}
\usepackage{graphicx}
\usepackage{cite,color}
\usepackage{stmaryrd}
\usepackage{algpseudocode}
\usepackage{caption}
\usepackage[ruled,linesnumbered]{algorithm2e}
\usepackage{amsmath,amssymb,amsfonts}
\newcommand{\RemoveAlgoNumber}{\renewcommand{\fnum@algocf}{\AlCapSty{\AlCapFnt\algorithmcfname}}}

\begin{document}

\title{A Novel Angle-Delay-Doppler Estimation Scheme for AFDM-ISAC System in Mixed Near-field and Far-field Scenarios}

\author{Yirui Luo,~\IEEEmembership{Student Member,~IEEE}, Yong Liang Guan,~\IEEEmembership{Senior Member,~IEEE}, Yao Ge,~\IEEEmembership{Member,~IEEE},

David~Gonz\'{a}lez~G.,~\IEEEmembership{Senior Member,~IEEE,} 
and Chau Yuen,~\IEEEmembership{Fellow,~IEEE}
        % <-this % stops a space
        \thanks{This work was supported under the RIE2020 Industry Alignment Fund—Industry Collaboration Projects (IAF-ICP) Funding Initiative, as well as cash and in-kind contribution from the industry partner(s). The work of Chau Yuen was supported by National Research Foundation FCP-NTU-RG-2024-025.}
        \thanks{Yirui Luo, Yong Liang Guan and Chau Yuen are with School of Electrical and Electronic Engineering, Nanyang Technological University, Singapore (e-mail: yirui001@e.ntu.edu.sg, eylguan@ntu.edu.sg, and chau.yuen@ntu.edu.sg.)}
        \thanks{Yao Ge is with the Continental-NTU Corporate Lab, Nanyang Technological University, Singapore (e-mail: yao.ge@ntu.edu.sg).}% <-this % stops a space
        \thanks{David~Gonz\'{a}lez~G. is with the Wireless Communications Technologies Group, Continental AG, Germany (e-mail: david.gonzalez.g@ieee.org).}
}



\maketitle
\begin{abstract}
The recently proposed multi-chirp waveform, affine frequency division multiplexing (AFDM), is considered as a potential candidate for integrated sensing and communication (ISAC). However, acquiring accurate target sensing parameter information becomes challenging due to fractional delay and Doppler shift occurrence, as well as effects introduced by the coexistence of near-field (NF) and far-field (FF) targets associated with large-scale antenna systems.
In this paper, we propose a novel angle-delay-Doppler estimation scheme for AFDM-ISAC system in mixed NF and FF scenarios. Specifically, we model the received ISAC signals as a third-order tensor that admits a low-rank CANDECOMP/PARAFAC (CP) format. By employing the Vandermonde nature of the factor matrix and the spatial smoothing technique, we develop a structured CP decomposition method that guarantees the condition for uniqueness. We further propose a low-complexity estimation scheme to acquire target sensing parameters with fractional values, including angle of arrival/departure (AoA/AoD), delay and Doppler shift accurately. We also derive the Cramér-Rao Lower Bound (CRLB) as a benchmark and analyze the complexity of our proposed scheme. Finally, simulation results are provided to demonstrate the effectiveness and superiority of our proposed scheme.

\end{abstract}


\section{Introduction}
The sixth-generation (6G) wireless systems are anticipated to maintain reliable communication in high mobility scenarios while significantly enhancing spectral and energy efficiencies, and supporting ubiquitous Internet of Things (IoT) connectivity for all devices\cite{Nguyen2022}. A variety of emerging 6G wireless services, such as unmanned vehicles, V2X, as well as the Industrial Internet of Things (IIoT), require wireless networks to provide prompt target sensing and communication capabilities \cite{Nguyen2022,Z_Zhang2019,lingsheng_10925171,liu}.  
Simultaneously, wireless communication and radar sensing systems are increasingly converging and expanding their frequency bands in response to the rising demand for high-speed communication and high-resolution radar sensing, leading to the spectrum congestion of the two systems. Therefore, integrated sensing and communication (ISAC) is regarded as a promising technology for realizing 6G wireless communications by integrating the two functions within a single system to improve spectrum and energy efficiency\cite{liu}.

In fact, ISAC has garnered significant attention from both academia and industry \cite{liu}. Most existing studies aim to achieve the integration of the two systems by employing the same waveform and hardware. In particular, orthogonal frequency division multiplexing (OFDM) is widely employed in ISAC due to the high spectral efficiency and robustness for target sensing \cite{10012421}. However, as the mobility and carrier frequency continues to increase in the 6G system, the resulting severe Doppler spreads can disrupt the orthogonality between subcarriers, thereby making OFDM suffer from intercarrier interference (ICI). One solution is to reduce the duration of the OFDM symbols to render the channel quasi-stationary during each symbol \cite{1638663}, but at a cost of lower spectral efficiency due to the increased cyclic prefix (CP) number. Another approach is to mitigate ICI \cite{ICI_weighted}, which, however, incurs the cost of a more complex transceiver design and is only effective at relatively low Doppler shift. Therefore, the development of new waveforms for the 6G wireless system to cope with doubly dispersive channels in high-mobility scenarios is particularly critical.

So far, several new modulation waveforms have been developed for doubly dispersive channels. The authors in \cite{R_Hadani} proposed a two-dimensional modulation technique named orthogonal time frequency space (OTFS). Unlike traditional OFDM in the time-frequency (TF) domain, OTFS symbols are modulated in the delay-Doppler (DD) domain, thereby achieving potential full diversity as each data symbol spans to the entire TF domain \cite{9508932,10891132,9738478}. Furthermore, this modulation allows to directly interact with channel delay and Doppler shift parameters, satisfying the sensing requirements of the ISAC system \cite{10791452}. Another kind of modulation waveform to cope with doubly dispersive channels is the chirp-based signals whose frequency varies linearly with time. Orthogonal chirp division multiplexing (OCDM) \cite{OCDM} exemplifies this approach and has been demonstrated to outperform OFDM in doubly dispersive channels due to the unique features of the chirp spread spectrum \cite{OCDM_performance}. However, the diversity order of OCDM in the doubly dispersive system cannot be optimal due to the limited representation of the channel delay-Doppler profile.

Recently, affine frequency division multiplexing (AFDM), a new multi-chirp waveform based on one-dimensional discrete affine Fourier transform (DAFT), is proposed in \cite{AFDM_TWC}. By adapting the chirp parameters of the DAFT based on the maximum channel delay and Doppler shift information, AFDM can separate all the paths in the discrete affine Fourier (DAF) domain and let each symbol experience all path coefficients, thus obtaining full diversity in the doubly dispersive system \cite{AFDM_TWC,luoqu,10711268,yiwei}. Meanwhile, compared to OTFS, it has been proved that AFDM has comparable BER performance but with less pilot overhead \cite{AFDM_TWC,luoqu,10711268,W_Benzine_CS,AFDM_MIMO} and lower implementation complexity \cite{K_R_R_Ranasinghe}. Moreover, due to the ability to support reliable communication in high-mobility scenarios and to distinguish multiple targets in the DAF domain, AFDM is considered as a promising candidate for 6G ISAC system.

Until now, the research of AFDM-based ISAC is still in its infancy. Authors in \cite{bistatic_AFDM} study a bistatic static target sensing scenario by using AFDM. Unfortunately, their method is unavailable for velocity sensing. For high-mobility target sensing, a simple match filter-based method for ISAC was proposed in \cite{Y_Ni_ISAC}, while the estimation performance of the Doppler shift is limited by the number of AFDM symbols in use, which is undesirable due to the requirement of low latency for next generation wireless networks. Furthermore, the match filter-based method requires an assumption that the delay is on the grid, which may be impractical in real wireless systems because of the limited spectral and temporal resources, resulting in insufficient resolution for both delay and Doppler shift. In order to obtain super-resolution estimations of both delay and Doppler shift, the authors in \cite{ML} proposed an approximate maximum likelihood (ML) algorithm by using just one AFDM symbol. However, ML suffers from a high computational complexity in an exhausted 2D grid search. Authors in \cite{K_R_R_Ranasinghe} and \cite{AFDM_ICC} exploited the sparsity of delay and Doppler in the DAF domain and proposed Bayesian principle based approaches for sensing parameter estimation. Although the methods in \cite{K_R_R_Ranasinghe} and \cite{AFDM_ICC} have the computational cost advantage over that in \cite{ML}, they still suffer from a high complexity due to the inverse of a large matrix. These aforementioned issues inspire us to develop a low-complexity algorithm that can estimate fractional delay and fractional Doppler shift, thereby enhancing the sensing accuracy.

Multiple-input multiple-output (MIMO) technique can provide additional spatial degrees of freedom (DoF) \cite{Z_Xiao}, which helps to enable spatial multiplexing, spatial diversity, and angle estimation. So far, MIMO technique has been widely employed in both communication and sensing systems. Authors in \cite{two_stage} proposed a two-stage estimation algorithm where the angle of arrival (AoA) is estimated by multiple signal classification (MUSIC), in which the range and velocity of the targets are matched by the estimation AoA. To avoid searching, authors in \cite{ESPRIT} proposed a super-resolution parameter estimation method via rotational invariance techniques (ESPRIT) for MIMO-OFDM ISAC system. Besides, based on the angle sparsity, authors in \cite{Z_Gao} \cite{J_Lee} proposed the compressed sensing (CS)-based methods for AoA, AoD, and delay estimation. Recently, tensor-based signal processing has received wide attention in the fields of channel estimation \cite{low_rank,Y_Lin,ruoyu}, radar parameter estimation \cite{D_Nion} \cite{M_Cao}, ISAC \cite{tensor_OTFS,ruoyu_ISAC,yirui}, etc., due to the ability to exploit the multidimensional characteristics of MIMO channels for parameters matching and performance enhancement. 

However, all the aforementioned works only consider far-field (FF) scenarios and may not be directly applicable to near-field (NF) scenarios. {As the MIMO array size continues to increase, the Rayleigh distance also grows, allowing for a larger number of targets to potentially fall within the near-field region \cite{near_dai} \cite{L_Wei}. This trend will significantly change the electromagnetic (EM) properties of the wireless environment from planar-wave propagation to spherical-wave propagation, leading to an inevitable near-field effect \cite{10149471}.} The NF-ISAC framework was first investigated by the authors in \cite{10135096} with a full-duplex antenna for simultaneous transmission and reception. To reduce power consumption and construction cost, authors in \cite{10579914} proposed a novel framework with double-array structure for NF integrated sensing, positioning and communication. To deal with the mixed NF and FF scenarios, authors in \cite{W_Zuo} proposed a subspace-based method and authors in \cite{8753714} proposed a fourth-order
cumulant base method for source localization. However, the aforementioned studies do not consider the case with high mobility.

Against the background, we propose a novel angle-delay-Doppler estimation scheme for AFDM-ISAC system in the mixed NF and FF scenarios in this paper. 
The main contributions of our work are summarized as follows:

\begin{itemize}
\item[1)]
We propose a novel AFDM-ISAC scheme for parameter estimation of mixed NF and FF targets, which takes into account the AoA, AoD, delay, and Doppler shift simultaneously. Specifically, we formulate the received AFDM-ISAC signal as a third-order tensor that admits a low-rank CANDECOMP/PARAFAC (CP) format, whose factor matrices contain the multi-target parameters. The high-dimensional estimation problem is parameterized by a small number of physical parameters due to the sparsity of the system.
\end{itemize}
 
\begin{itemize}
\item[2)]
 We apply a structured CP decomposition method by exploiting the Vandermonde nature of the factor matrix and the spatial smoothing technique. We also analyze the uniqueness condition for the CP decomposition, which defines the feasible applications of our proposed scheme, and demonstrate that the condition can be relaxed.
\end{itemize}

\begin{itemize}
\item[3)]
We develop a subspace-based method that avoids high-complexity eigendecomposition for AoA estimation with mixed NF and FF targets. We further propose a super-resolution joint delay-Doppler estimation scheme by leveraging the full representation of the channel delay-Doppler profile provided by AFDM in the DAF domain, where the fractional components can be estimated by using one-dimensional iterative process for sensing enhancement with low complexity.
\end{itemize}
 
\begin{itemize}
\item[4)]
 Finally, we derivate the Cramér-Rao lower bound (CRLB) for the corresponding parameter estimation and evaluate the performance of our proposed AFDM-ISAC scheme with NF and FF targets in terms of the normalized mean square error (NMSE) of the estimated AoA, AoD, delay and Doppler shift. Simulation results indicate the performance of the proposed scheme outperforms the other benchmark schemes and is also close to that of the corresponding CRLB.
\end{itemize}

The remainder of the paper is organized as follows. Section II depicts the signal model of the AFDM-based mixed NF and FF ISAC system. Section III explains the establishment of the structured tensor model and analyzes the uniqueness of CP decomposition. Section IV introduces our proposed algorithm for AoA, AoD, delay, and Doppler shift estimation. Section V provides the CRLB and complexity analysis of our scheme. Section VI presents the numerical results, followed by concluding remarks in Section VII.

Notation: If not specified, lowercase letter $x$, lowercase bold letter ${\mathbf{x}}$, and uppercase bold letter ${\mathbf{X}}$ donate variables, vectors, and matrices, respectively. The calligraphy letter $\boldsymbol{\mathcal{X}}$ represents tensor and its mode-n unfolding matrix is donated by ${{\mathbf{X}}_{\left( n \right)}}$. ${\left(  \cdot  \right)^{ - 1}}$, ${\left(  \cdot  \right)^{\dag}}$, ${\left(  \cdot  \right)^{ *}}$, ${\left(\cdot\right)^{T}}$, ${\left(\cdot\right)^{-T}}$ and ${\left(\cdot\right)^{H}}$ stands for the inverse, pseudo-inverse, conjugate, transpose, transpose-inverse and conjugate transpose, respectively. $ \circ $, $ \odot $, and $* $ stand for the outer product, the Khatri-Rao and Hadamard product, respectively. Let $\mathbf{X}\in \mathbb{C}^{M\times N}$, $\overline{\mathbf{X}}=\mathbf{X}\left(2:M,:\right)$ stands for the matrices without the first row of $\mathbf{X}$. The identity matrix and all-zeros matrix are denoted by $\mathbf{I}_N\in \mathbb{C}^{N\times N}$ and $\mathbf{0}_{M\times N}\in \mathbb{C}^{M\times N}$, respectively.

\section{Mixed NF and FF AFDM-ISAC System Model}
\begin{figure}
\centering
\includegraphics[width=8cm]{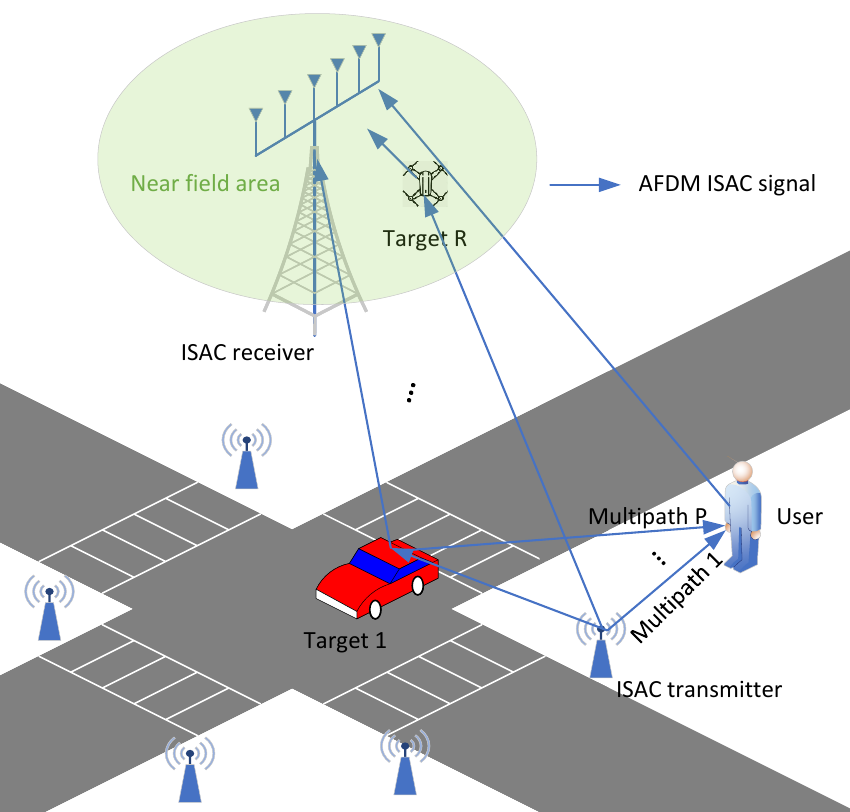}
\caption{{Mixed near-field and far-field AFDM-ISAC system model.}}
\label{ISAC}
\end{figure}

As illustrated in Fig. \ref{ISAC}, we consider a AFDM-ISAC system with the coexistence of NF and FF targets operating over a channel with a carrier frequency $f_c$ and bandwidth $BW$. {Specifically, the micro-cell base station, equipped with a uniform linear array (ULA) of $K$ antennas, transmits the AFDM ISAC signals to communicate with a specific communication user and serves as the ISAC transmitter. The AFDM ISAC signals can also be reflected by environmental targets and subsequently reused by a nearby macro-cell base station for sensing purpose.} The macro-cell base station serves as the ISAC receiver and is equipped with a ULA consisting of $G$ antennas. Considering the size of the macro-cell base station array is much larger than that of the micro-cell base station, we assume a one-side mixed NF and FF sensing channel model for the ISAC system. Specifically, we make an FF assumption at the ISAC transmitter side and a mixed NF and FF assumption at the ISAC receiver side.
Moreover, the signals transmitted by the ISAC transmitter (i.e., micro-cell base station) are known by the ISAC receiver (i.e., macro-cell base station) due to the backhaul connection facilitated by optical fiber.
\begin{figure*}
\centering
\includegraphics[width=16cm]{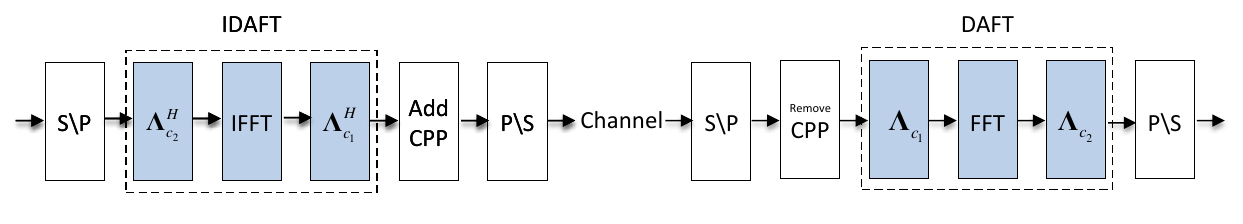}
\caption{Block diagram for AFDM transmitter and receiver.}
\label{processing}
\end{figure*}
\subsection{Transmitted AFDM ISAC Signal}
In this subsection, we provide a concise overview of transmitted AFDM signal as introduced in \cite{AFDM_TWC}. The ISAC transmitter and receiver processing diagrams of the AFDM system are shown in Fig. \ref{processing}. 
Consider an AFDM symbol with a bandwidth $BW=N_c\Delta f$ and a duration of $T_a$, where $N_c$ is the number of the chirp subcarriers and $\Delta f=1/T_a$ is the chirp subcarrier spacing. Let $\mathbf{x}$ be a $N_c \times 1$ symbol vector in the DAF domain. To transform $\mathbf{x}$ into the time domain, an $N_c$-point inverse DAFT (IDAFT) is applied, yielding

\begin{equation}
    s\left[ n \right] = \sum\limits_{m = 0}^{N_c - 1} {x\left[ m \right]{\phi _n}\left( m \right)},\label{s[n]}
\end{equation}
where $n=0,\dots,N_c-1$ and ${\phi _n}\left( m \right) = \frac{1}{{\sqrt N_c }}\exp \left[ {j2\pi \left( {{c_1}{n^2} + {c_2}{m^2} + nm/N_c} \right)} \right]$. As defined in \cite{AFDM_TWC}, $c_1>0$ and $c_2>0$ are the AFDM chirp parameters to be optimized for full diversity. In the matrix form,  \eqref{s[n]} can be expressed as ${\mathbf{s}} = {\boldsymbol{\Lambda }}_{{c_1}}^H\mathbf{F}^H{\boldsymbol{\Lambda }}_{{c_2}}^H{\mathbf{x}}$, where ${{\mathbf{\Lambda }}_{c_i}} = {\text{diag}}\left( {{e^{ - j2\pi c_i{n^2}}},n = 0,1,...,N_c - 1,i=1,2} \right)$ and $\mathbf{F}$ is the normalized $N_c$-point DFT matrix.
Before transmitting, a chirp-periodic prefix (CPP) of length $L_{cpp}$ should be appended to let the channel lie in a periodic domain, which has an expression of 
\begin{equation}
    s[n] = s\left[ {N_c + n} \right]{{\text{e}}^{ - j2\pi {c_1}\left( {{N_c^2} + 2N_cn} \right)}},{\text{     }}n =  - {L_{cpp}},..., - 1,
\end{equation}
where $L_{cpp}$ is set to be an integer equal to or larger than the maximum channel delay taps. We can also write down $s\left(t\right)$ as a continuous version of the transmitted AFDM signal with the expression of 
\begin{equation}
    s\left( t \right) = \frac{1}{{\sqrt T_a }}\sum\limits_{m = 0}^{N_c - 1} {x\left[ m \right]} {{\text{e}}^{j2\pi \left( c_2m^2+\Phi_m\left(t\right)\right)}},{\text{  }}0 \leqslant t < T_a,
\end{equation}
where $\Phi_m\left(t\right)$ is defined in segments over intervals that correspond to the partition $\{t_{m,q}\}_{q=0,...,2N_cc_1}$ of $[0,T_a)$ with $t_{m,0}=0$ and $t_{m,q}=\frac{N_c-m}{2Nc_1}T_s+\frac{q-1}{2c_1}T_s$ for $q>0$. $T_s=T_a/N_c$ is the sampling interval and $\Phi_m\left(t\right)$ is defined as $\Phi_m\left(t\right)=\frac{{{c_1}}}{{T_s^2}}{t^2} + \frac{m}{{N_c{T_s}}}t-\frac{q}{T_s}t$ for $t\in\left[t_{m,q},t_{m,q+1}\right)$ \cite{ML}.

\subsection{Sensing Channel Model}

\begin{figure}
\centering
\subfigure[]{\includegraphics[height=1in]{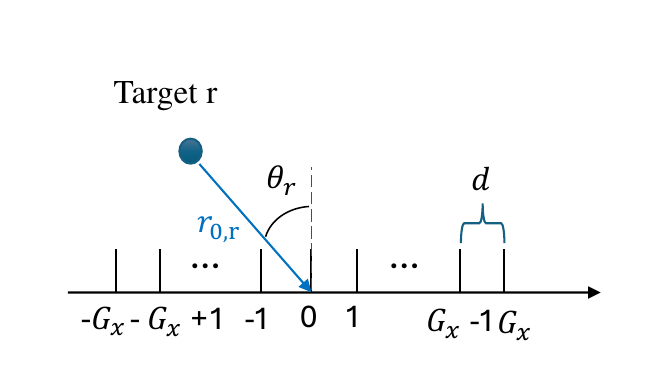}}\hspace{-0.1in}
\subfigure[]{\includegraphics[height=1in]{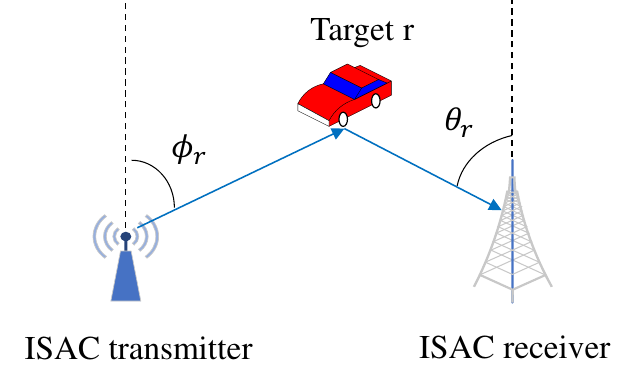}}
\caption{(a) ULA configuration of ISAC receiver. (b) Illustration of system geometry.}
\label{angle_figure}
\end{figure}
Consider a doubly dispersive sensing channel with $R$ targets for sensing purpose, and each target is only characterized by its line-of-sight (LOS) path. 
As depicted in Fig. \ref{angle_figure} (a), we assume the ISAC receiver has a symmetrical ULA consisting of $G=2G_x+1$ antennas along $x$-axis, where the middle element is located at the origin and set to be the reference antenna.

{The array response of the ISAC receiver for target $r$ at the $g$-th antenna of the ISAC receiver is given by \cite{W_Zuo}}
\begin{equation}
\begin{split}
   { \left[{{\mathbf{a}}_R}\left( {\theta_r ,r_{0,r}} \right)\right]_g}&
   {= {e^{j\frac{{2\pi }}{\lambda }\left( {\sqrt {r_{0,r}^2 + {{\left( {gd} \right)}^2} - 2{r_{0,r}}gd\sin \theta_r  - {r_{0,r}}} } \right)\sin \theta_r }}}\\
    & {\approx  {e^{ jg\rho_r  + jg^2\xi_r }},}\label{near_aoa}
\end{split}
\end{equation}
{where $\left[{{\mathbf{a}}_R}\left( {\theta_r ,r_{0,r}} \right)\right]_g$ is the $g$-th entry of array response vector ${{\mathbf{a}}_R}\left( {\theta_r ,r_{0,r}} \right)$, $\rho_r=-\frac{{2\pi d}}{\lambda }\sin \theta_r $ and $\xi_r=\frac{{\pi {d^2}}}{{\lambda r_{0,r}^2}}\cos \theta_r$, for $g=-G_x,-G_x+1,...,G_x$. $d$ is the receive antenna spacing, $\lambda$ is the wavelength, $\theta_r$ is the AoA of the $r$-th target and $r_{0,r}\in \left[0.62{\left( {{{{D^3}} / \lambda }} \right)^{1/2}}, +\infty \right)$ is the distance between the $r$-th target and the reference antenna (i.e., $g=0$). When $r_{0,r}$ is within the Rayleigh distance (i.e., $r_{0,r}\leq 2D^2/\lambda$), the phase of the electromagnetic wave from near-field target $r$ impinging on the ISAC receiver has to be calculated based on the spherical wave model and the phase shift caused by distance information cannot be ignored.} 

If the target $r$ is in the far field, i.e., $r_{0,r}\in \left(2D^2/\lambda, + \infty \right)$, $jg^2\xi_r$ is approximated equal to zero and \eqref{near_aoa} simplifies to
\begin{equation}
    \left[{{\mathbf{a}}_R}\left( {\theta_r } \right)\right]_g= e^{ jg\rho_r}.\label{far_aoa}
\end{equation}

{Even though the array response of the ISAC receiver for the far-field targets can be simplified to \eqref{far_aoa}, the form of \eqref{near_aoa} is employed throughout to provide a unified representation for both near-field and far-field targets, thereby simplifying the overall expression.}
Denote the scattering coefficient, total range, and relative velocity of target $r$ as $\gamma_r$, $L_r$, and $v_r$, respectively. Then, the total delay and Doppler shift can be expressed as $\tau_{r}=\frac{L_r}{c}$ and $f_{d,r}=\frac{v_r f_c}{c}$ with $c$ denotes the speed of light. Thus, the matrix impulse response for the time-varying MIMO sensing channel can be expressed as
\begin{equation}
    {\mathbf{H}}\left( {t,\tau } \right) = {\sum\limits_{r = 1}^{R}  {{\gamma_r}{\mathbf{a}}_R\left( {{\theta _r}},r_{0,r} \right){\mathbf{a}}_T\left( {{\phi _r}} \right)} ^H}\delta \left( {\tau  - {\tau _r}} \right){e^{j2\pi {f_{d,r}}t}},\label{channel}
\end{equation}
where 
\begin{equation}
      {\mathbf{a}_T}\left( \phi_r  \right) = {\left[1,{{e^{ - j\pi \sin \phi_r }}},...,{{e^{ - j\pi\left( {K - 1} \right)\sin \phi_r }}} \right]^T}\label{AOD_steer}
 \end{equation}
 is the linear array response of the ISAC transmitter with half wavelength antenna spacing and $\phi_r$ denotes the AoD of target $r$. The system geometry explanation for AoA and AoD is illustrated in Fig.~\ref{angle_figure} (b).

\subsection{Received AFDM ISAC Signals}

After through the channel in \eqref{channel}, the received time domain signal at the $g$-th antenna of the ISAC receiver can be expressed as
\begin{equation}
\begin{split}
    {r}_g\left( t \right) =&\sum\limits_{r = 1}^{R} \sum\limits_{k = 0}^{K-1}{{{\gamma_r}\left[{\mathbf{a}}_R\left( {{\theta _r}},r_{0,r} \right)\right]_g{\left[ {{\mathbf{a}}_T\left( {{\phi _r}} \right)} \right]}_k} {e^{j2\pi {f_{d,r}}t}}}\\
    &\times {s}\left( {t-\tau_{r}} \right)  + {w}_g\left( t \right), 
\end{split} 
\end{equation}
 where $w_g\left(t\right)$ is the additive Gaussian noise in the time domain received by the $g$-th antenna with $g=-G_x,...,G_x$ and ${\left[ {{\mathbf{a}}_T\left( {{\phi _r}} \right)} \right]}_k$ is the $k$-th entry of the vector ${ {{\mathbf{a}}_T\left( {{\phi _r}} \right)} }$. By sampling with the interval $T_s$, the received samples $\mathbf{R} \left[ n \right]\in {\mathbb{C}^{G \times 1}}$ in the time domain at time instant $n$ can be expressed as
\begin{equation}
\begin{split}
    \mathbf{R}\left[ n \right] =& \sum\limits_{r = 1}^{R} \sum\limits_{k = 0}^{K-1} {{\gamma_r}{\mathbf{a}}_R\left( {{\theta _r}},r_{0,r} \right){{{\left[ {{\mathbf{a}}_T\left( {{\phi _r}} \right)} \right]}_k}}}{e^{j2\pi {\nu_r}n/N_c}}\\
    &\times s\left[ {n - \beta_r} \right] + \mathbf{W}\left[n \right], \label{receive_time}
\end{split}
\end{equation}
with $n = 0, \ldots ,\left( {N_c + {L_{cpp}}} \right) - 1$. The normalized delay and Doppler shift of target $r$ are given by $\beta_r=\tau_r/T_s$ and $\nu_r=N_cf_{d,r}T_s$. We define $\beta_r=\ell_r+\iota_r$ and $\nu_r=\alpha_r+\mu_r$ with $\ell_r \in \left(0,\ell_{\max}\right]$, $\iota_r\in \left( { - \frac{1}{2}} \right.,\left. {\frac{1}{2}} \right]$, $\alpha_r \in \left[-\alpha_{\max},\alpha_{\max}\right]$ and $\mu_r\in \left( { - \frac{1}{2}} \right.,\left. {\frac{1}{2}} \right]$ referring to the integer normalized delay, fractional normalized delay, integer normalized Doppler shift, and fractional normalized Doppler shift, respectively, where $\ell_{\max}$ is the maximum normalized delay and $\alpha_{\max}$ is the maximum normalized Doppler shift in the ISAC system.

 To avoid overlapping for different non-zero entries at the same position, $c_1$ is set to be \cite{AFDM_TWC}
 \begin{equation}
     c_1=\frac{{2\left( {{\alpha _{\max }} + {k_v}} \right) + 1}}{{2N_c}},
 \end{equation}
where $k_v$ is the parameter according to the Doppler spread. Moreover, the maximum normalized delay and Doppler shift should satisfy
\begin{equation}
    2\left({\alpha_{\max }}+k_v\right) + {\ell_{\max }} + 2\left({\alpha_{\max }}+k_v\right){\ell_{\max }} < N_c.
\end{equation}
to achieve the optimal diversity order.

 After serial to parallel conversion and discarding CPP, the received time domain sensing signal is transferred back to the DAF domain through DAFT. The received signal in the DAF domain can be expressed as 
  \begin{equation}
     \mathbf{Y}\left[ {m} \right] = \sum\limits_{n = 0}^{N_c - 1} \mathbf{R} \left[ {n} \right]\phi _n^*\left( m \right),     
\end{equation}
with $m=0,...,N_c-1$.
In the matrix form, the output ${\mathbf{Y}} \in \mathbb{C}^{G\times N_c}$ can be represented by

\begin{equation}
\begin{split}
    {\mathbf{Y}} = \sum\limits_{r = 1}^{R} \sum\limits_{k = 0}^{K-1}{{\gamma_r}}{\mathbf{a}}_R\left( {{\theta _r}},r_{0,r} \right){{{\left[ {{\mathbf{a}}_T\left( {{\phi _r}} \right)} \right]}_k}}{\left(\mathbf{E}_r\mathbf{x}\right)}^T + {\mathbf{\tilde W}},\label{DAF-receive}
\end{split}
\end{equation}
where 
\begin{equation}
    \mathbf{E}_r={{\mathbf{\Lambda }}_{{c_2}}}{\mathbf{F}}{{\mathbf{\Lambda }}_{{c_1}}}{\mathbf{\Gamma }}_{CP{P_r}}{{\mathbf{\Delta }}_{{\nu_r}}}{{\mathbf{\Pi }}^{\beta_r}}{\mathbf{\Lambda }}_{{c_1}}^H{{\mathbf{F}}^H}{\mathbf{\Lambda }}_{{c_2}}^H,\label{Er}
\end{equation}
and ${\mathbf{\tilde W}} = {{\mathbf{\Lambda }}_{{c_2}}}{\mathbf{F}}{{\mathbf{\Lambda }}_{{c_1}}}{\mathbf{W}}$ is the corresponding noise in the DAF domain. ${{\mathbf{\Delta }}_{{\nu_r}}}$ has an expression of ${{\mathbf{\Delta }}_{{\nu_r}}} = {\text{diag}}({e^{ - j2\pi {{\left( {0:{N_c} - 1} \right){\nu_r}} /{N_c}}}})$ and ${\mathbf{\Pi }}^{\beta_r}$ has the expression of ${\mathbf{\Pi }}^{\beta_r}={{\mathbf{F}}^H}{\text{diag}}({e^{ - j2\pi {\left( {0:{N_c} - 1} \right) \beta_r}/ {N_c}}}){\mathbf{F}}$. ${{\boldsymbol{\Gamma }}_{CP{P_r}}}$ is a diagonal matrix with the expression of
 \begin{equation}\label{CPP}
\begin{split}
{{\boldsymbol{\Gamma }}_{CP{P_r}}} = {\text{diag}}\left( {\left\{ {\begin{array}{*{20}{c}}
  {{{{e}}^{ - j2\pi {c_1}\left( {{N_c^2} - 2N_c\left( {{\ell_r} - n} \right)} \right)}}}&{n < {\ell_r}} \\ 
  1&{n \geqslant {\ell_r}} 
\end{array}} \right.} \right).
\end{split}
\end{equation}
 We can see from \eqref{CPP} that if $N_c$ is an even number and $2N_cc_1$ is an integer, $\boldsymbol{\Gamma}_{CPP_r}=\mathbf{I}$.

Meanwhile, the AFDM signal received at the communication user side can be expressed as
 \begin{equation}
 \begin{split}
     {\mathbf{y}_{com}} &= \sum\limits_{p = 1}^{P}\sum\limits_{k = 0}^{K-1}\gamma_p {{{\left[ {{\mathbf{a}}_T\left( {{\phi _p}} \right)} \right]}_k}} {{\mathbf{\Lambda }}_{{c_2}}}{\mathbf{F}}{{\mathbf{\Lambda }}_{{c_1}}}{{\mathbf{\Gamma }}_{CP{P_p}}}\\
     &\times{{\mathbf{\Delta }}_{{\nu_p}}}{{\mathbf{\Pi }}^{\beta_p}}{\mathbf{\Lambda }}_{{c_1}}^H{{\mathbf{F}}^H}{\mathbf{\Lambda }}_{{c_2}}^H{\mathbf{x}} + {\mathbf{\tilde W}},
 \end{split}
 \end{equation}
 where $P$ is the number of paths and $\gamma_p$ is the pathloss of the $p$-th path. In this paper, we mainly focus on the target sensing algorithm design. Details for the communication processing, i.e., channel estimation and symbol detection, can refer to the reference\cite{AFDM_TWC}. 
\section{Tensor Representation And Analysis}
In this section, we first elucidate how to formulate the received sensing signal into a tensor model. We then discuss the uniqueness of the structured CP decomposition.
\subsection{Low-rank Tensor Model}
By utilizing \eqref{DAF-receive} with its structure, we can represent the received DAF domain signal as a third-order tensor $\boldsymbol{\mathcal{Y}} \in {\mathbb{C}^{{G} \times {N_c} \times {K}}}$, which is given by
\begin{equation}
    \boldsymbol{\mathcal{Y}} = \boldsymbol{\mathcal{X}} + \boldsymbol{\mathcal{W}},\label{CPD}
\end{equation}
where $\boldsymbol{\mathcal{X}} \in {\mathbb{C}^{{G} \times {N_c} \times {K}}}$ is a tensor form for the received signal in the DAF domain and $\boldsymbol{\mathcal{W}}\in {\mathbb{C}^{{G} \times {N_c} \times {K}}}$ is the corresponding noise in the DAF domain. 

%\textit{Definition 1 (Rank-one Tensor)}: An $M$-th order tensor $\boldsymbol{\mathcal{X}} \in {\mathbb{C}^{{I_1} \times {I_2} \times  \cdots  \times {I_M}}}$ is rank-one if it can be expressed as an outer product of $M$ vectors.

%\textit{Definition 2 (CP decomposition)}: If a tensor can be construed as a weighted sum of rank-one outer products, its decomposition can be represented in the CANDECOMP/PARAFAC (CP) decomposition form.

It is worth noting that $\boldsymbol{\mathcal{X}}$ can be written in the CANDECOMP/PARAFAC (CP) decomposition form, i.e.,
\begin{equation}
\begin{split}
    \boldsymbol{\mathcal{X}} &= \left[\kern-0.15em\left[ {{\boldsymbol{\gamma }};{{\mathbf{A}}_R},{{\mathbf{B}}_C},{{\mathbf{A}}_T}} 
 \right]\kern-0.15em\right] \\
 &= \sum\limits_r {{\gamma _r}{\mathbf{a}}_R\left( {{\theta _r},r_{0,r}} \right)}  \circ {\mathbf{b}}\left( {{\beta_r},{\nu_r}} \right) \circ {\mathbf{a}}_T\left( \phi_r \right),
 \end{split}
\end{equation}
where ${\boldsymbol{\gamma }}=\left[\gamma_1,\gamma_2,...,\gamma_R\right]^T$ and
\begin{equation}
    {\mathbf{b}}\left( {{\beta_r},{\nu_r}}\right)={{\mathbf{\Lambda }}_{{c_2}}}{\mathbf{F}}{{\mathbf{\Lambda }}_{{c_1}}}{{\mathbf{\Gamma }}_{{CPP}_{r}}}{{\mathbf{\Delta }}_{ {\nu_r}}}{{\mathbf{\Pi }}^{\beta_r }}{\mathbf{\Lambda }}_{{c_1}}^H{{\mathbf{F}}^H}{\mathbf{\Lambda }}_{{c_2}}^H{\mathbf{x}}.\label{br}
\end{equation}
%with $\beta_r=\ell_r+\iota_r$.

Additionally, ${{\mathbf{A}}_R} \in \mathbb{C}^{G\times R}$, ${{\mathbf{B}}_C}\in \mathbb{C}^{N_c\times R}$ and ${{\mathbf{A}}_T}\in \mathbb{C}^{K\times R}$ are the related 
% rank-$R$
factor matrices with the respective expression as
\[\begin{gathered}
  {{\mathbf{A}}_{\mathbf{R}}} = \left[ {\begin{array}{*{20}{c}}
  {{\mathbf{a}}_R\left( {{\theta _1},r_{0,1}} \right)}&{{\mathbf{a}}_R\left( {{\theta _2},r_{0,2}} \right)}& \cdots &{{\mathbf{a}}_R\left( {{\theta _R},r_{0,R}} \right)} 
\end{array}} \right], \hfill \\
  {{\mathbf{B}}_{\mathbf{C}}} = \left[ {\begin{array}{*{20}{c}}
  {{\mathbf{b}}\left( {{\beta_1},{\nu_1}} \right)}&{{\mathbf{b}}\left( {{\beta_2},{\nu_2}} \right)}& \cdots &{{\mathbf{b}}\left( {{\beta_R},{\nu_R}} \right)} 
\end{array}} \right], \hfill \\
  {{\mathbf{A}}_{\mathbf{T}}} = \left[ {\begin{array}{*{20}{c}}
  {{\mathbf{a}}_T\left( {{\phi _1}} \right)}&{{\mathbf{a}}_T\left( {{\phi _2}} \right)}& \cdots &{{\mathbf{a}}_T\left( {{\phi _R}} \right)}
\end{array}} \right]. \hfill \\ 
\end{gathered} \]

\subsection{Uniqueness Property Analysis}
The uniqueness condition of the CP decomposition problem is crucial for target parameter estimation, which can ensure that the decomposed factor matrices contain accurate information of targets' parameters. According to Kruskal’s condition \cite{Kruskal}, the uniqueness of the CP decomposition can be guaranteed by 
\begin{equation}
    {k_{{{\mathbf{A}}_R}}} + {k_{{{\mathbf{B}}_C}}} + {k_{{{\mathbf{A}}_T}}} \geqslant 2R + 2,\label{k-condition}
\end{equation}
where $ {k_\mathbf{A}}$ is the Kruskal rank of matrix $\mathbf{A}$ and $R$ is the rank of the tensor. In the generic case\footnotemark{}, the condition \eqref{k-condition} becomes
\begin{equation}
    \min (G,R) + \min (N_c,R) + \min (K,R) \geqslant 2R + 2.\label{con11}
\end{equation}
\footnotetext{Generic case means that the CP decomposition property holds with probability one when the entries of factor matrices are drawn from absolutely continuous probability density functions.}

The condition \eqref{k-condition} specifies that each factor matrix is required to possess a Kruskal-rank greater than one, which implies that there are no linearly dependent columns in one factor matrix and the system must include at least two targets.
%\textit{Definition 3 (Vandermonde Matrix)}: The factor matrix ${\mathbf{A}}$ is defined as a Vandermonde matrix if its columns satisfy
%\[{\mathbf{a}}_{r} = {\left[ {1,{z_{r}},z_{r}^2, \ldots ,z_{r}^{{I} - 1}} \right]^T}\]
%for $r=1,...,R$. And $\left\{ {{z_{r}}} \right\}_{r = 1}^{R}$ are called the generators of ${\mathbf{A}}$.
By leveraging the Vandermonde nature of ${\mathbf{A}}_T$ due to the inherent structures of far-field ULA and spatial smoothing, the unique condition of the CP decomposition of $\boldsymbol{\mathcal{X}}$ can be relaxed to \cite{van_cpd} 
\begin{equation}
\left\{ \begin{gathered}
  \text{rank}\left( {{\mathbf{A}}_T^{\left( {{K_3} - 1} \right)} \odot {{\mathbf{A}}_R}} \right) = R, \hfill \\
  \text{rank}\left( {{\mathbf{A}}_T^{\left( {{L_3}} \right)} \odot {{\mathbf{B}}_C}} \right) = R, \hfill \\ 
\end{gathered}  \right.\label{con}
\end{equation}
where ${K_3} + {L_3} = {K} + 1$ and $K_3 \in \left[1,K\right]$ is the parameters for spatial smoothing \cite{van_cpd}. ${{\mathbf{A}}^{\left( M \right)}}$ denotes the sub-matrix which contains the first $M$ rows of matrix $\mathbf{A}$. Generically, the unique condition \eqref{con} is satisfied if
\begin{equation}
   \min \left( {\left( {{K_3} - 1} \right)G,{L_3}{N_c}} \right)  \geqslant  R.\label{con13}
\end{equation}

With the structural constraints, the condition \eqref{con13} can be satisfied when $R=1$ and even holds for the case that $\mathbf{A}_R$ or $\mathbf{B}_C$ is not full Kruskal-rank (i.e., some targets have the same AoA or delay and Doppler shift).

\section{Proposed Angle-Delay-Doppler Estimation Scheme}
In this section, we propose a novel tensor-based angle-delay-Doppler estimation scheme for AFDM-ISAC system with mixed NF and FF targets in two phases. In the first phase, we exploit the structural characteristics of the factor matrices in our model to perform CP decomposition. In the second phase, we extract the relevant interested sensing parameters from the estimated factor matrices obtained in the first phase.
\subsection{Structured CP Decomposition for Factor Matrices}
Before CP decomposition, the number of the targets needs to be known and can be easily obtained by the minimum description length (MDL) approach \cite{995060}. Then, the CP decomposition of \eqref{CPD} can be achieved by solving
\begin{equation}
    \mathop {\min }\limits_{{{{\mathbf{\hat A}}}_R},{{{\mathbf{\hat B}}}_C}{\mathbf{,}}{{{\mathbf{\hat A}}}_T}} \left\| {\boldsymbol{\mathcal{Y}} - \sum\limits_r {{{\hat \gamma }_r}{\mathbf{\hat a}_R}\left( {{\theta _r},r_{0,r}} \right)}  \circ {\mathbf{\hat b}}\left( {{\beta_r},{\nu_r}} \right) \circ {\mathbf{\hat a}_T}\left( {{\phi_r}} \right)} \right\|_F^2.\label{CPD_min}
\end{equation}

One of the most renowned algorithms for solving \eqref{CPD_min} is the alternating least squares (ALS) scheme, which alternatively optimizes one of the factor matrices while keeping the other two factor matrices fixed to minimize the data fitting error until convergence \cite{tensor}. However, the outcome of the alternating process is significantly influenced by the initial state, potentially leading to convergence towards a biased local optimum. Additionally, ALS treats factor matrices as random ones rather than structured ones, which consequently impacts performance and effectiveness due to the neglect of prior information. More importantly, the computational complexity is sometimes high because of the slow convergence speed of this iterative method.

To address the aforementioned issues, we develop an enhanced structured tensor decomposition approach in this paper. To be specific, we take the unfolding matrix of $\mathcal{Y}$ along its second dimension and we have
\begin{equation}
    {\mathbf{Y}}_{\left( 2 \right)} = {{\mathbf{B}}_C}\left( {{{\mathbf{A}}_T} \odot {{\mathbf{A}}_R}} \right)^T + {\mathbf{W}}_{\left( 2 \right)},
\end{equation}
and ${\mathbf{W}}_{\left( 2 \right)} \in \mathbb{C}^{N_c\times KG}$ is the unfolding matrix of $\boldsymbol{\mathcal{W}}$ along its second dimension. Due to the Vandermonde nature of ${\mathbf{A}}_T$, the spatial smoothing technique can be applied to ${\mathbf{Y}}_{\left( 2 \right)}$ for expanding the dimension and also overcoming the problems for the rank deficient case. %expanding the dimension of ${\mathbf{Y}}_{\left( 2 \right)}$ and also overcoming the problems for the rank deficient case, 

Specifically, we define an integer pair $\left( {{K_3},{L_3}} \right)$ with ${K_3} + {L_3} = {K} + 1$ and a cyclic selection matrix ${{\mathbf{J}}_{l_3}} \in {\mathbb{C}^{{K_3}{G} \times {KG}}}$ with the expression of
\begin{equation}
{{\mathbf{J}}_{{l_3}}} \triangleq \left[ {\begin{array}{*{20}{c}}
  {{{\mathbf{0}}_{{K_3} \times \left( {{l_3} - 1} \right)}}}&{{{\mathbf{I}}_{{K_3}}}}&{{{\mathbf{0}}_{{K_3} \times \left( {{L_3} - {l_3}} \right)}}} 
\end{array}} \right] \otimes {{\mathbf{I}}_G},
\end{equation}
where $ {l_3} \in \left[1,L_3\right]$. Then, the processing matrix after spatial smoothing can be obtained by
\begin{equation}
\begin{split}
    {\boldsymbol{\Upsilon}} &=\left[ {\begin{array}{*{20}{c}}
  {{{\mathbf{J}}_1}{\mathbf{Y}}_{\left( 2 \right)}^T}&{{{\mathbf{J}}_2}{\mathbf{Y}}_{\left( 2 \right)}^T}&{...}&{{{\mathbf{J}}_{{L_3}}}{\mathbf{Y}}_{\left( 2 \right)}^T} 
\end{array}} \right]+\mathbf{W}_{\Upsilon} \\
    &=\left( {{\mathbf{A}}_T^{\left( {{K_3}} \right)} \odot {{\mathbf{A}}_R}} \right){\mathbf{\Lambda }}{\left( {{\mathbf{A}}_T^{\left( {{L_3}} \right)} \odot {{\mathbf{B}}_C}} \right)^T}+\mathbf{W}_{\Upsilon},\label{C_near}
\end{split}
\end{equation}
with ${\mathbf{\Lambda }} = {\text{diag}}\left( {{{ {\gamma _1}, \ldots ,{\gamma _R}} }} \right)$. $\mathbf{W}_{\Upsilon} \in \mathbb{C}^{K_3G\times L_3N_c}$ is the corresponding noise after spatial smoothing.

According to \cite{van_cpd}, we utilize the ESPRIT-like approach for the factor matrices estimation. In particular, we compute the singular value decomposition (SVD) $\boldsymbol{\Upsilon} = {\mathbf{US}}{{\mathbf{V}}^H}$, where ${{\mathbf{U}}} \in {\mathbb{C}^{K_3G \times R}}$ and ${{\mathbf{V}}} \in {\mathbb{C}^{L_3N_c \times R}}$ are singular matrices. $\mathbf{S}\in \mathbb{C}^{R\times R}$ is a diagonal matrix formed by $R$ singulars. Then, there exists a nonsingular matrix $\mathbf{M} \in \mathbb{C}^{R\times R}$ to satisfy 
\begin{equation}
    \mathbf{UM}=\mathbf{A}_T^{\left(K_3\right)}\odot {{\mathbf{A}}_R},\label{UM}
\end{equation}
and
\begin{equation}
    {{\mathbf{V}}^*}{\mathbf{S }}{{\mathbf{P}}}=\mathbf{A}^{\left(L_3\right)}_T\odot {{\mathbf{B}}_C}.\label{VSP}
\end{equation}
where ${\mathbf{P}} = {{\mathbf{M}}^{ - T}}$.
Define the generators of $\mathbf{A}_T$ as $\left\{ {{z_{r}}} \right\}_{r = 1}^{R}$. Due to the Vandermonde structure of $\mathbf{A}_T$, we have
\begin{equation}
    \left( {{{\mathbf{A}}_T^{\left(K_3-1\right)}} \odot {{\mathbf{A}}_R}} \right){\mathbf{Z}} = {\overline{\mathbf{A}}_T^{\left(K_3\right)}} \odot {{\mathbf{A}}_R},
\end{equation}
with $\mathbf{Z}=\text{diag}\left(z_1,z_2,...,z_R\right)$. Then, the relation \eqref{UM} can also be expressed as 
\begin{equation}
    \mathbf{U}_1\mathbf{M}= \left( {{{\mathbf{A}}_T^{\left(K_3-1\right)}} \odot {{\mathbf{A}}_R}} \right),\text{ } \mathbf{U}_2\mathbf{M}= {\overline{\mathbf{A}}_T^{\left(K_3\right)}} \odot {{\mathbf{A}}_R},
\end{equation}
where ${{\mathbf{U}}_1} = {{\mathbf{U}}}\left( {1:\left( {{K_3} - 1} \right)G,:} \right)$ and ${{\mathbf{U}}_2} = {{\mathbf{U}}}\left( {\left( {G + 1} \right):{K_3}G,:} \right)$. $\mathbf{M}$ and $\mathbf{Z}$ can be obtained by the eigenvalue decomposition (EVD) ${\mathbf{U}}_1^\dag {{\mathbf{U}}_2} = {\mathbf{MZ}}{{\mathbf{M}}^{ - 1}}$, and the columns of the estimated $\hat{\mathbf{A}}_T$ can be generated by
\begin{equation}
    \hat{\mathbf{a}}_{T}\left(\phi_r\right)={\left[ {1,{\hat{z}_{r}},\hat{z}_{r}^2, \ldots ,\hat{z}_{r}^{K - 1}} \right]^T},\label{AK}
\end{equation}
where ${{\hat z}_{r}}=\frac{Z\left[  r,r\right]}{{\left| {Z\left[  r,r\right]} \right|}}$ is the estimated generator of the $r$-th column of $\hat{\mathbf{A}}_T$ for $r=1,..,R$, and $Z\left[  r,r\right]$ is the $\left(r,r\right)$-th entry of $\mathbf{Z}$.

According to \eqref{UM} and \eqref{AK}, the columns of the estimated $\hat{\mathbf{A}}_R$ can be obtained by
\begin{equation}
    \hat{\mathbf{a}}_{R}\left(\theta_r,r_{0,r}\right)=\left( {\left(\hat{\mathbf{a}}_{T,r}^{\left( {{K_3}} \right)}\right)^H \otimes {{\mathbf{I}}_{G}}} \right){\mathbf{U}}{{\mathbf{m}}_r},\label{AG_near}
\end{equation}
for $r=1,...,R$. ${{\mathbf{m}}_r}$ is the $r$-th column of $\mathbf{M}$ and $\hat{\mathbf{a}}_{T,r}$ is the simplified expression of $\hat{\mathbf{a}}_{T}\left(\phi_r\right)$.

Finally, from \eqref{VSP}, the columns of $\hat{\mathbf{B}}_C$ can be obtained by
\begin{equation}
   \hat {\mathbf{b}}_{r} = \left( {\frac{{\left({\mathbf{\hat a}}_{T,r}^{\left( {{L_3}} \right)}\right)^H}}{{{\left({\mathbf{\hat a}}_{T,r}^{\left( {{L_3}} \right)}\right)^H}{\mathbf{\hat a}}_{T,r}^{\left( {{L_3}} \right)}}} \otimes {{\mathbf{I}}_{{N_c}}}} \right){{\mathbf{V}}^*}{\mathbf{\Sigma }}{{\mathbf{p}}_r},\label{br}
\end{equation} 
where ${{\mathbf{p}}_r}$ is the $r$-th column of $\mathbf{P}$ and $r=1,...,R$. $\hat{\mathbf{b}}_{r}$ is the simplified expression of $\hat{\mathbf{b}}\left(\beta_r,\nu_r\right)$.

The structured CP decomposition algorithm is summarized in \textbf{Algorithm 1}.

\begin{algorithm}
    \caption{Structured CP Decomposition}
    \KwIn{${\mathbf{Y}}_{\left( 2 \right)}$, number of the targets $R$}
    Select an integer pair $\left( {{K_3},{L_3}} \right)$ subject to ${K_3} + {L_3} = K + 1$\;
    Perform spatial smoothing for ${\mathbf{Y}}_{\left( 2 \right)}$ according to \eqref{C_near}\; %If all the targets are in the far field, we can also perform \eqref{C_FAR}\;
    Reconstruct $\hat{\mathbf{A}}_T$ by \eqref{AK}\;
    Reconstruct $\hat{\mathbf{A}}_R$ by \eqref{AG_near}\;% or \eqref{AG_far} for total far field case\;
    Obtain $\hat{\mathbf{B}}_C$ according to \eqref{br}\;
    \KwOut{$\hat{\mathbf{A}}_R$, $\hat{\mathbf{B}}_C$, $\hat{\mathbf{A}}_T$.}
\end{algorithm}

\subsection{Sensing Parameters Estimation}
This subsection explains how we extract the relevant interested sensing parameters
from the estimated factor matrices obtained in Algorithm 1.
\subsubsection{AoD}
Each column of ${{{\mathbf{\hat A}}}_T}$ is associated with one AoD. Since we already got the generators of ${{{\mathbf{\hat A}}}_T}$, the AoD can be estimated directly by 
\begin{equation}
    {\hat{\phi} _r} = \arcsin \left( {\frac{{\angle {\hat{z}_{r}}}}{-\pi }} \right),\label{phi}
\end{equation}
where $\angle$ is the phase angle extraction operator and $r=1,...,R$.

\subsubsection{AoA}
The AoA information of target $r$ can be extracted from the covariance matrix $\mathbf{C}_{R,r}$ with the expression of 
\begin{equation} \mathbf{C}_{R,r}=\hat{\mathbf{a}}_R\left(\theta_r,r_{0,r}\right)\hat{\mathbf{a}}_R\left(\theta_r,r_{0,r}\right)^H
\end{equation}

To resolve the coupling between distance and AOA caused by the spherical wavefront of the near-field targets, we form an $\left(G_x+1\right)\times \left(G_x+1\right)$ Toeplitz correlation matrix $ {\mathbf{T}_{R,r}}$ with the expression of
\begin{equation}
\begin{split}
    {\mathbf{T}_{R,r}}& =\\
    &\! \left[ {\begin{array}{*{20}{c}}
  {{c_{R,r}}\left( {{G_x} + 1} \right)}&{{c_{R,r}}\left( {{G_x}} \right)}& \cdots &{{c_{R,r}}\left( 1 \right)} \\ 
  {{c_{R,r}}\left( {{G_x} + 2} \right)}&{{c_{R,r}}\left( {G_x + 1} \right)}& \cdots &{{c_{R,r}}\left( 2 \right)} \\ 
   \vdots & \vdots & \ddots & \vdots  \\ 
  {{c_{R,r}}\left( {2{G_x} + 1} \right)}&{{c_{R,r}}\left( {2{G_x}} \right)}& \cdots &{{c_{R,r}}\left( {{G_x} + 1} \right)} 
\end{array}} \right] \\
    &=\! {{\tilde {\mathbf{a}}}_R}\left( {{\theta _r}} \right){{\tilde  {\mathbf{a}}}_R}^H\left( {{\theta _r}} \right),\label{CG}
\end{split}
\end{equation}
where $c_{R,r}\left(a\right)$ is the $\left(a,2G_x+2-a\right)$-th entry of $\mathbf{C}_{R,r}$, ${\mathbf{\tilde a}_R}\left( {{\theta _r}} \right) = {\left[ {1,{e^{ j2{\rho _r}}},....,{e^{j2G_x{\rho _r}}}} \right]^T}$ and $d$ is set to be a quarter wavelength to avoid ambiguity \cite{W_Zuo} for AoA estimation. By leveraging \eqref{CG}, the distance and AOA
information can be effectively decoupled{\footnotemark{}}.\footnotetext{Regardless of whether the target $r$ is a near-field or far-field target, we can obtain the Toeplitz correlation matrix in \eqref{CG}, which only contains the AoA information. Therefore, our proposed AoA estimation scheme does not require differentiation between near-field and far-field targets.}

For estimating AoA of target $r$ in \eqref{CG}, the multiple signal classification (MUSIC) algorithm is commonly used. However, the MUSIC algorithm suffers from a high complexity of eigendecomposition. To avoid the high complexity  eigendecomposition, we derive ${\mathbf{T}}_{R,r}$ to two parts as ${{\mathbf{T}}_{R,r}} = {\left[ {{\mathbf{T}}_{1,r}^T,{\mathbf{T}}_{2,r}^T} \right]^T}$, where ${\mathbf{T}}_{1,r}$ is the first row of ${\mathbf{T}}_{R,r}$ while ${\mathbf{T}}_{2,r}$ is the last $G_x$ rows of ${\mathbf{T}}_{R,r}$. Then, $\theta_r$ can be obtained by 
\begin{equation}
     \hat \theta_r  =\mathop {\min }\limits_\theta  {\mathbf{\tilde a}}_R^H\left( \theta  \right){{\mathbf{\Pi }}_{{\mathbf{Q}_r}}}{{\mathbf{\tilde a}}_R}\left( \theta  \right),\label{theta}
\end{equation}
where 
\begin{equation}
    {{\mathbf{\Pi }}_{{\mathbf{Q}_r}}} = {{\mathbf{Q}}_r}{\left( {{\mathbf{Q}}_r^H{{\mathbf{Q}}_r}} \right)^{ - 1}}{\mathbf{Q}}_r^H,
\end{equation}
\begin{equation}
    {{\mathbf{Q}}_r} = {\left[ {{\mathbf{P}}_r^T, - {{\mathbf{I}}_{G_x}}} \right]^T},
\end{equation}
and
\begin{equation}
   {{\mathbf{P}}_r} = {\left( {{{\mathbf{T}}_{1,r}}{\mathbf{T}}_{1,r}^H} \right)^{ - 1}}{{\mathbf{T}}_{1,r}}{\mathbf{T}}_{2,r}^H.
\end{equation}

\subsubsection{Delay and Doppler Shift}
Both the delay and Doppler shift information are included in $\hat{\mathbf{B}}_C$. However, a direct 2D exhaustive search requires high complexity. Thus, by leveraging the full representation of the channel delay-Doppler profile provided by AFDM in the DAF domain, we propose a two-stage super-resolution joint delay-Doppler estimation scheme, where fractional components can be estimated by using a one-dimensional iterative process for sensing enhancement with low complexity.

To enhance the understanding of the proposed scheme, we first present an intuitive representation of the delay-Doppler profile provided by AFDM in the DAF domain, as included in $\hat{\mathbf{b}}\left(\beta_r,\nu_r\right)$. Recall the vector representation in \eqref{br}, the $m$-th entry of $\mathbf{b}\left(\beta_r,\nu_r\right)$ can be expressed as
\begin{equation}
\begin{split}
    b_r\left[ m \right] =& \frac{1}{{{N_c}}}\sum\limits_{m' = 0}^{{N_c} - 1} {x\left[ {m'} \right]} {e^{j2\pi \left( -\frac{{{\beta _r}m'}}{{{N_c}}}+{{c_1}\beta _r^2 + {c_2}\left( {{m'^2} - m{^2}} \right) } \right)}}\\
    &\times \underbrace {\sum\limits_{n = 0}^{{N_c} - 1} {{e^{j\frac{{2\pi n }}{{{N_c}}}\left( {m' - m - lo{c_r}} \right)}}{e^{j2\pi {\iota _r}{\mathcal{C}_{m,n,r}}}}} }_{{\mathcal{F}_r}\left[ {m,m'} \right]}, \label{bm}
\end{split}
\end{equation}
where ${\mathbf{b}}_r$ is the simplified expression of ${\mathbf{b}}\left(\beta_r,\nu_r\right)$. $lo{c_r} = {\left( {2N_c{c_1}{\beta_r} - {\nu _r}} \right)_{N_c}}$, ${\mathcal{C}_{m,n,r}} = \sum\limits_{q = 0}^{2{N_c}{c_1}} {q{I_{{\mathcal{L}_{m,q}}}}\left( {{{\left( {n - {\beta _r}} \right)}_{{N_c}}}} \right)} $, and ${\left(  \cdot  \right)_{N_c}}$ is the modulo $N_c$ operation. ${I_{{\mathcal{L}_{m,q}}}}$ is the indicator function of the set ${\mathcal{L}_{m,q}}=\{n_{m,q}+1,...,n_{m,q+1}\}$, where $n_{m,0}=0$ and $n_{m,q}=\left\lfloor {\frac{{{N_c} - m}}{{2{N_c}{c_1}}} + \frac{{q - 1}}{{2{c_1}}}} \right\rfloor$ for $q>0$ \cite{ML}. 
\begin{algorithm}
% \SetNlSty{def}{(}{)}
% \SetAlgoLined
% \DontPrintSemicolon
%\small
\KwIn{Factor matrix $\hat{\mathbf{B}}_C$, transmit symbol vector $\mathbf{x}$}
 Initialization: $\eta =\frac{{\sqrt 5  - 1}}{2}$\;  
\For{$r = 1:R$}
{Obtain $\left( {{{\hat \ell}_r},{{\hat \alpha }_r}} \right)$ by \eqref{integer};\\
 $\mathbf{y}_r={{\mathbf{\Lambda }}_{{c_2}}^H}{\mathbf{F}^H}{{\mathbf{\Lambda }}_{{c_1}}^H}{{\hat {\mathbf{b}}}_r}$, ${l^{\left( {0} \right)}} = {{\hat \ell}_r},{\text{ }}{\alpha ^{\left( {0} \right)}} = {{\hat \alpha }_r}$, $t=1$\;
\For{$t \leq T$}
{${a_l} = {\alpha ^{\left( {t - 1} \right)}} - 1,{a_u} = {\alpha ^{\left( {t - 1} \right)}} + 1$;\\
\Repeat{Stopping criteria} 
{${g_1} = {a_u} - \eta \left( {{a_u} - {a_l}} \right),{g_2} = {a_l} + \eta \left( {{a_u} - {a_l}} \right)$;\\
${f_{1}} =  {\left| {\left( {{{\mathbf{\Delta }}_{g_1}}{{\mathbf{\Pi }}^{{l^{\left( {t - 1} \right)}}}}{\mathbf{s}}} \right)^H}{{\mathbf{y}}_r} \right|^2} $,\\${f_{2}} =  {\left|{\left( {{{\mathbf{\Delta }}_{g_2}}{{\mathbf{\Pi }}^{{l^{\left( {t - 1} \right)}}}}{\mathbf{s}}} \right)^H}{{\mathbf{y}}_r} \right|^2}$\;
\textbf{if} ${f_{1}} \leq {f_{2}}$, ${a_l} = {g_1},{g_1} = {g_2}$\;
\textbf{if} ${f_{1}} > {f_{2}}$, ${a_u} = {g_2},{g_2} = {g_1}$\;}
${\alpha ^{\left( t \right)}} = \frac{{{a_l} + {a_u}}}{2}$\;
${b_l} = {l^{\left( {t - 1} \right)}} - 1,{b_u} = {l^{\left( {t - 1} \right)}} + 1$\;
\Repeat{Stopping criteria} 
{${q_1} = {b_u} - \eta \left( {{b_u} - {b_l}} \right),{q_2} = {b_l} + \eta \left( {{b_u} - {b_l}} \right)$;\\
${f_{3}} =  {\left| {\left( {{{\mathbf{\Delta }}_{\alpha^{\left(t\right)}}}{{\mathbf{\Pi }}^{q_1}}{\mathbf{s}}} \right)^H}{{\mathbf{y}}_r} \right|^2}$,\\${f_{4}} =  {\left| {\left( {{{\mathbf{\Delta }}_{{\alpha^{\left(t\right)}}}}{{\mathbf{\Pi }}^{q_2}}{\mathbf{s}}} \right)^H}{{\mathbf{y}}_r} \right|^2}$\;
\textbf{if} ${f_{3}} \leq {f_{4}}$, ${b_l} = {q_1},{q_1} = {q_2}$\;
\textbf{if} ${f_{3}} > {f_{4}}$, ${b_u} = {q_2},{q_2} = {q_1}$\;}
${l ^{\left( t \right)}} = \frac{{{b_l} + {b_u}}}{2}$\;
$t=t+1$\;}
${{\hat \beta}_r} = {l^{\left( T \right)}}$, ${{\hat \nu }_r} = {\alpha ^{\left( T \right)}}$\;
$\hat \tau_r={{\hat \beta}_r} t_s$, $\hat f_{d,r}=\nu_r \Delta f$\;}
\KwOut{Delay $\hat{\boldsymbol{\tau}}$, and Doppler shift $\hat{\mathbf{f}}_{d}$ of the targets.}
\caption{Proposed Two-stage Delay and Doppler Shift Estimation Scheme}
\end{algorithm}

Since $c_2$ is set to be a number sufficiently smaller than $\frac{1}{{2N_c}}$ \cite{AFDM_TWC}, the value of ${N_c{c_2}\left( {m'^2 - {m^2}} \right)}$ is approximately approach zero, which can be ignored. 
It is worth noting that ${{\mathcal{F}_r}\left[ {m,m'} \right]}$ in \eqref{bm} simplifies to $\frac{{{e^{  j2\pi \left( { m'-m- lo{c_r}} \right)}} - 1}}{{{e^{ j\frac{{2\pi }}{{{N_c}}}\left( {m' - m - lo{c_r}} \right)}} - 1}}$ for integer normalized Doppler shift case, and to $N_c\delta \left( {m' - m - lo{c_r}} \right)$ when both the normalized delay and Doppler shift have zero fractional components.

\begin{figure}
\centering
\subfigure[Signal vector before pulse compression.]{\includegraphics[width=2.8in]{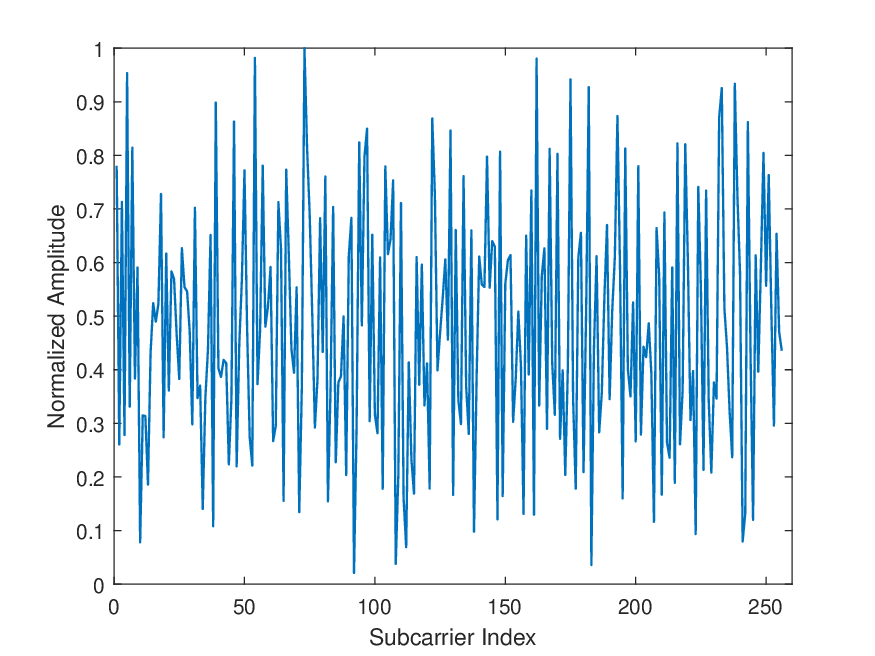}}\hspace{-0.1in}
\subfigure[Signal vector after pulse compression.]{\includegraphics[width=2.8in]{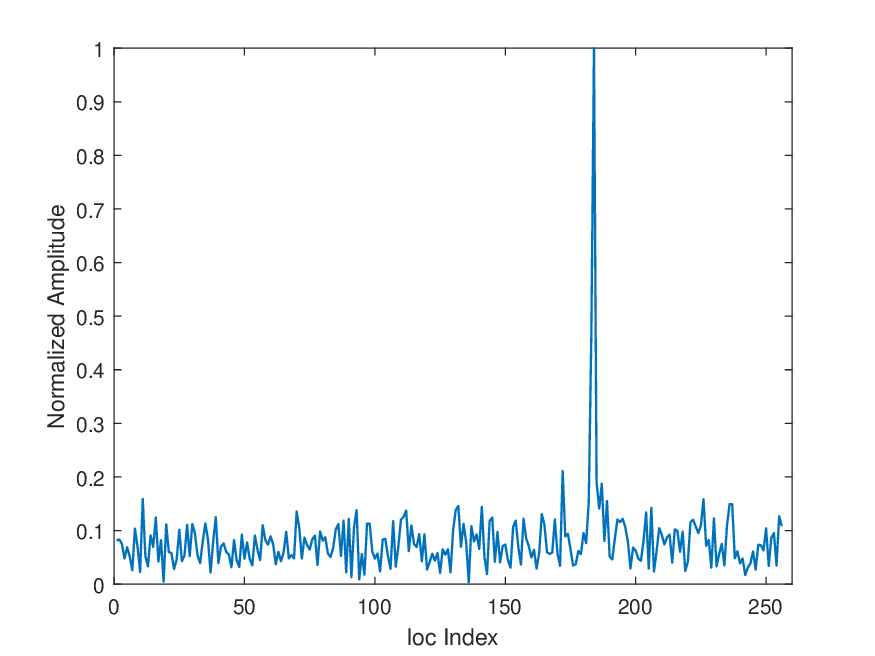}}
\caption{The result comparison of 1D pulse compression performance.}
\label{pulse_com}
\end{figure}

Inspired by pulse compression in radar sensing, we propose a low-complexity estimation scheme based on one-dimentional (1D) pulse compression in stage one according to \eqref{bm}. Specifically, the vector for target $r$ after 1D pulse compression is donated as ${\mathbf{u}}_{r}$, with the $p$-th element expressed as
\begin{equation}
    {u_r}\left[ p \right] = \sum\limits_{m = 0}^{{N_c} - 1} {{{\hat b}_r}^ * \left[ m \right]} x\left[ {{{\left( {m - p} \right)}_{{N_c}}}} \right]{e^{ - j\frac{{2\pi }}{{{N_c}}}\left\lfloor {\frac{{{N_c} + 1 - p}}{{2{N_c}{c_1}}}} \right\rfloor {{\left( {m-p} \right)}_{{N_c}}}}},\label{pulse}
\end{equation}
 where $p\in \left[0,N_c-1\right]$ and $r=1,...,R$. To provide a better illustration, Fig.~\ref{pulse_com} shows an example of our proposed method for one target with $\beta=8.13$, $\nu=1.67$, $c_1=\frac{9}{512}$ and $c_2=0$ at $\text{SNR}=15$ dB. In this example, information symbols in $\mathbf{x}$ are randomly generated 16-QAM symbols, and the subcarrier number $N_c$ is set to be $256$. The DAF domain vector ${\mathbf{b}}_r$ is shown in Fig.~\ref{pulse_com} (a), and the vector ${\mathbf{u}}_r$ after 1D pulse compression is shown in Fig.~\ref{pulse_com} (b). The results in Fig.~\ref{pulse_com} clearly demonstrate that after the pulse compression, the vector $\mathbf{u}_r$ contains the information of interest about this target with the noise. The value of $loc_r$ can be obtained directly from the index of the peak in $\mathbf{u}_r$.

{Due to the predefined $c_1$, the DAF domain impulse response of our AFDM system can constitute a full delay-Doppler representation of the sensing channel \cite{AFDM_TWC}. Therefore, the delay and Doppler shift information can be obtained unambiguously from ${\hat {loc{_r}}}$ by}
\begin{equation}
    {{{\hat \ell}_r} = \left\lfloor {\frac{\hat {loc{_r}}}{{2N_cc_1}}} \right\rfloor ,  \quad {{\hat \alpha }_r}=  2N_cc_1{{\hat \ell}_r}-{\hat {loc{_r}}}.}\label{integer}
\end{equation}

 However, the pulse compression-based method can only obtain the integer components of the normalized delay and Doppler shift. Therefore, in stage two, we need to perform an off-grid search based on the integer components we obtained in the first stage. To further reduce the complexity, we employ an iterative 1D golden section search method for the estimation of fractional parts of normalized delay and Doppler shift.
The two-stage low-complexity delay and Doppler shift estimation scheme is summarized in \textbf{Algorithm 2}.

\begin{algorithm}
    \caption{Proposed Angle-Delay-Doppler Estimation Scheme}
    \KwIn{${\mathbf{Y}}_{\left( 2 \right)}$, number of the targets $R$ }
    \textbf{For} $r=1$ to $R$ \textbf{do}\\
    Perform structured CP decomposition according to  \textbf{Algorithm 1}\;
    Estimate $\hat \phi_r$ according to \eqref{phi}\;
    Estimate $\hat \theta_r$ according to \eqref{theta}\;
    Calculate $\hat{\tau}_r$ and $\hat{f}_{d,r}$ via \textbf{Algorithm 2}\;
    \textbf{End}\;
    \KwOut{Estimated AoA $\hat{\boldsymbol{\theta}}$, AoD $\hat{\boldsymbol{\phi}}$, delay $\hat{\boldsymbol{\tau}}$, and the Doppler shift $\hat{\mathbf{f}}_{d}$ of the targets.}
\end{algorithm}

The locations and the relative velocities for all the targets can be obtained after the estimation of parameters of interest. The whole processing of our proposed angle-delay-Doppler estimation scheme is summarized in \textbf{Algorithm 3}.

\section{Performance Analysis}
\subsection{CRLB} 
In this subsection, we develop the Cramér-Rao lower bound (CRLB) for the target parameter estimation. Note that the CRLB serves as a minimum limit for the variance of any unbiased estimator \cite{CRLB}. It also acts as a benchmark for evaluating the effectiveness of the proposed approach. 
The CRLB for the estimated parameters can be calculated by the inverse of the Fisher information matrix (FIM) \cite{CRLB_cpd}
\begin{equation}    {\text{CRB}}\left(\boldsymbol{\Psi}\right) = {\boldsymbol{\Omega} ^{ - 1}}\left( \boldsymbol{\Psi} \right),\label{CRLB}
\end{equation}
where the estimated parameters $\boldsymbol{\Psi}$ includes $\boldsymbol{\Psi}=\left[\boldsymbol{\theta},{\mathbf{r}_0},\boldsymbol{\tau},{\mathbf{f}_d},\boldsymbol{\phi},\boldsymbol{\gamma }\right]$, with
${\boldsymbol{\theta }} = {\left[ {{\theta _1},...,{\theta _R}} \right]}$, ${\mathbf{r}_0} = {\left[ {{r_{0,1}},...,{r_{0,R}}} \right]}$, ${\boldsymbol{\tau }} = {\left[ {{\tau _1},...,{\tau _R}} \right]}$, ${\mathbf{f}_d} = {\left[ {{f_{d,1}},...,{f_{d,R}}} \right]}$, ${\boldsymbol{\phi }} = {\left[ {{\phi _1},...,{\phi _R}} \right]}$ and ${\boldsymbol{\gamma }} = {\left[ {{\gamma _1},...,{\gamma _R}} \right]}$.

{Here, it is worth mentioning that although the proposed scheme does not require to estimate the values of $\gamma_r$ and $r_{0,r}$, they still influence the CRLB. Consequently, we have to include them as estimated variables in the derivation calculations.}
Details for the derivation can be found in the Appendix. 
\subsection{Computational Complexity}
The computational complexity of our proposed angle-delay-Doppler estimation scheme for AFDM-ISAC system in mixed NF and FF scenarios is elaborated as follows. The truncated SVD of ${\boldsymbol{\Upsilon}}$ has a computational complexity of $\mathcal{O}\left(K_3L_3GN_cR\right)$ and the EVD of ${\mathbf{U}}_1^\dag {{\mathbf{U}}_2}$ has the complexity of $\mathcal{O}\left(R^2K_3G+R^3\right)$. The complexity of estimating the factor matrices $\mathbf{A}_R$, $\mathbf{B}_C$ and $\mathbf{A}_T$ are $\mathcal{O}\left(K_3GR^2\right)$, $\mathcal{O}\left( L_3N_cR^2\right)$ and $\mathcal{O}\left(KR \right)$, respectively.
For the complexity of parameters estimation, $\left\{ {{\phi_r}} \right\}_{r = 1}^R$ and $\left\{ {{\theta_r}} \right\}_{r = 1}^R$ has the complexity order of $\mathcal{O}\left(R \right)$ and $\mathcal{O}\left(RG\varepsilon_\theta \right)$, where $\varepsilon_\theta$ is the number of one dimensional AoA search grids. The complexity for the estimation of $\left\{ {{\tau_r}} \right\}_{r = 1}^R$ and $\left\{ {{f_{d,r}}} \right\}_{r = 1}^R$ are both $\mathcal{O}\left( R{N_c}\log {N_c}+TR{N_c}\log {N_c}\right)$. 
Thus, the total complexity of our proposed estimation scheme is  $\mathcal{O}\left(R{N_c}\left( {{K_3}{L_3}G + {L_3}R + 2\left( {T + 1} \right)\log {N_c}} \right)\right)$ $+\mathcal{O} \left({R^2}{K_3}\left( {2G + R} \right)\right)$ $ + \mathcal{O}\left(R\left( {K + G{\varepsilon _\theta } + 1} \right)\right)$.%The most expensive step is SVD, so the total complexity can be considered as $\mathcal{O}\left(F_3D_3GN_cR\right)$.
\section{Numerical Results}

In this section,  we conduct numerical simulations to investigate the performance of our proposed angle-delay-Doppler estimation scheme for mixed NF and FF AFDM-ISAC system based on Monte Carlo simulations. We consider a scenario where the micro-cell base station employs a ULA with $K= 8$ antenna elements, and the macro-cell base station employs a ULA with $G_x= 50$. {The spatial smoothing parameter $K_3$ is set to be $5$ and $L_3=4$.} The carrier frequency is 60 GHz. {So the Rayleigh distance in our simulations is $6.25$ m.} The number of subcarriers in the simulation is $N_c=256$ and we employ 16-QAM symbols. 
%In our simulations, the ISAC transmitter is located at $\left[0,0\right]$ (m), while the middle of the receiver's ULA has the coordination of $\left[400,0\right]$ (m).
The AoDs and AoAs are randomly distributed in $\left[-\pi/2,\pi/2\right]$, and the number of mixed NF and FF targets is set to be $R=3$ with one target in the near-field and two in the far-field. The maximum integer part of the normalized Doppler shift is considered to be $\alpha_{max}=1$, which corresponds to a maximum speed of $540$ km/h. The maximum delay is considered to be $\ell_{max}=12$, then $L_{cpp}$ is set $12$ to avoid ISI. Chirp parameters $c_1=\frac{9}{512}$ and $c_2=0$. The signal-to-noise ratio (SNR) is defined as
$$\mathrm{SNR} \triangleq \frac{{\left\| {\boldsymbol{\mathcal{Y}} - \boldsymbol{\mathcal{W}}} \right\|_F^2}}{{\left\| \boldsymbol{\mathcal{W}} \right\|_F^2}},$$
where $\boldsymbol{\mathcal{Y}}$ and $\boldsymbol{\mathcal{W}}$ represent the received signal and the  noise defined in \eqref{CPD}, respectively.

\begin{figure}
\centering
\includegraphics[width=8cm]{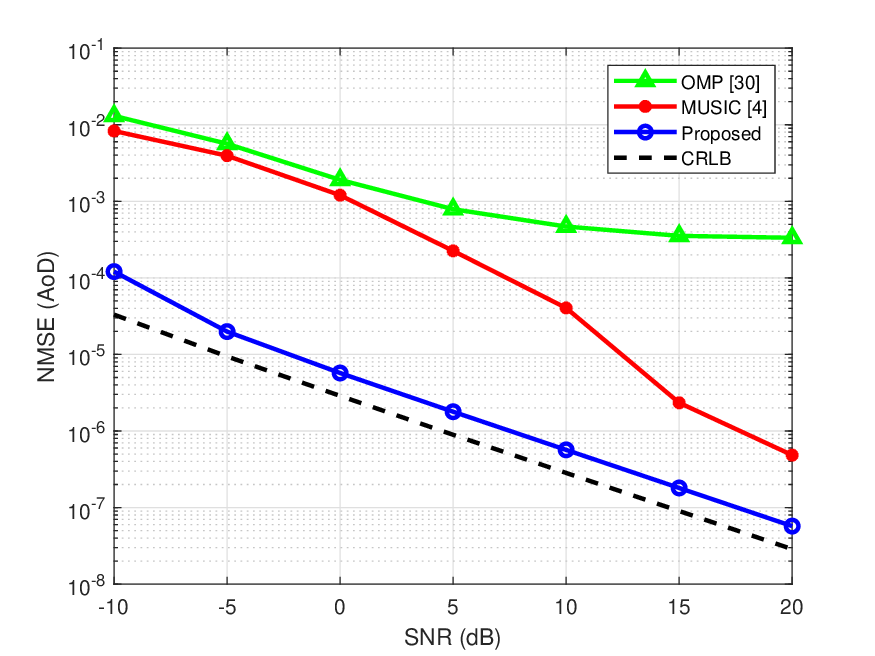}
\caption{The NMSE performance of AoD estimation versus SNR.}
\label{AOD}
\end{figure}

\begin{figure}
\centering
\includegraphics[width=8cm]{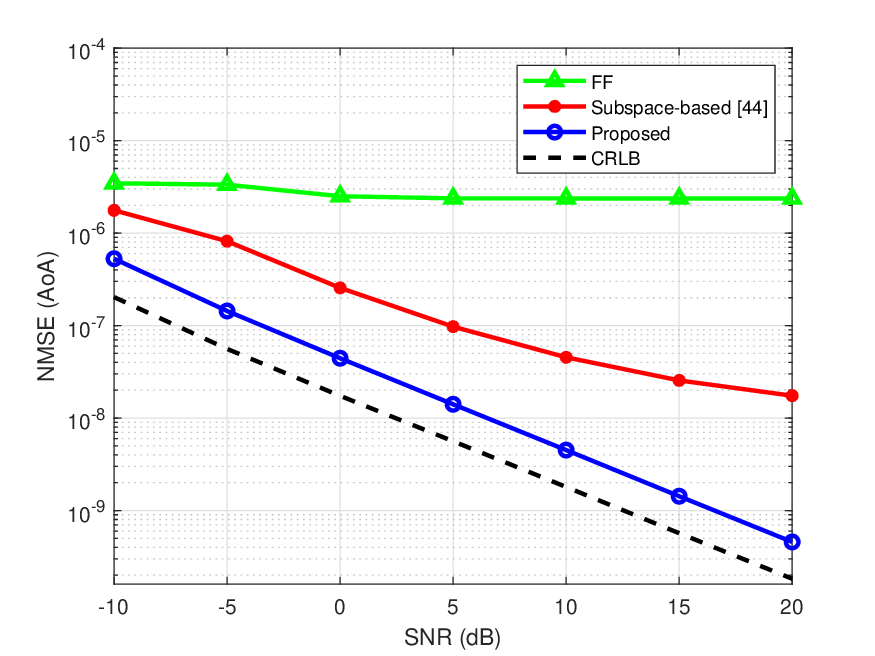}
\caption{The NMSE performance of AoA estimation versus SNR.}
\label{AOA}
\end{figure}

First, we evaluate the estimation accuracy of the target sensing parameters $\left\{\theta_r, \tau_r, f_{d,r},\phi_r \right\}_{r=1}^R$ measured by the normalized mean square error (NMSE), which are defined as 
$$\mathrm{NMSE}_{\theta} \triangleq \frac{{ {\left\| {{\boldsymbol{\theta}} - {{\hat {\boldsymbol{\theta}}}}} \right\|} _F^2}}{{ {\left\| {{\boldsymbol{\theta}}} \right\|} _F^2}}, \mathrm{NMSE}_{\tau} \triangleq \frac{{ {\left\| {{\boldsymbol{\tau}} - {{\hat {\boldsymbol{\tau}}}}} \right\|} _F^2}}{{ {\left\| {{\boldsymbol{\tau}}} \right\|} _F^2}},$$
$$\mathrm{NMSE}_{f_{d}} \triangleq \frac{{  {\left\| {{\mathbf{f}_{d}} - {{\hat {\mathbf{f}}}_{d}}} \right\|} _F^2}}{{ {\left\| {{\mathbf{f}_{d}}} \right\|} _F^2}}, \mathrm{NMSE}_{\phi} \triangleq \frac{{ {\left\| {{\boldsymbol{\phi}} - {{\hat {\boldsymbol{\phi}}}}} \right\|} _F^2}}{{ {\left\| {{\boldsymbol{\phi}}} \right\|} _F^2}}.$$

Fig.~\ref{AOD} depicts the NMSE performance of AoD versus SNR for our proposed scheme, which is compared with the MUSIC-based method \cite{liu}, orthogonal matching pursuit (OMP) method \cite{J_Lee} and the CRLB. The results demonstrate that the performance of our proposed scheme exhibits exponential improvement as the SNR increases. Also, the proposed scheme can achieve the lowest NMSE compared with the MUSIC-based method and OMP method, and is relatively close to the CRLB. The performance gain of our proposed method arises from the ability to capture the intrinsic multidimensional structure of the multi-way data and the gridless processing approach, which helps to be free from the grid discretization error. What's more, the performance of the MUSIC method is constrained by the number of transmitted antennas, and the OMP method is limited by the grid resolution.
\begin{figure}
\centering
\includegraphics[width=8cm]{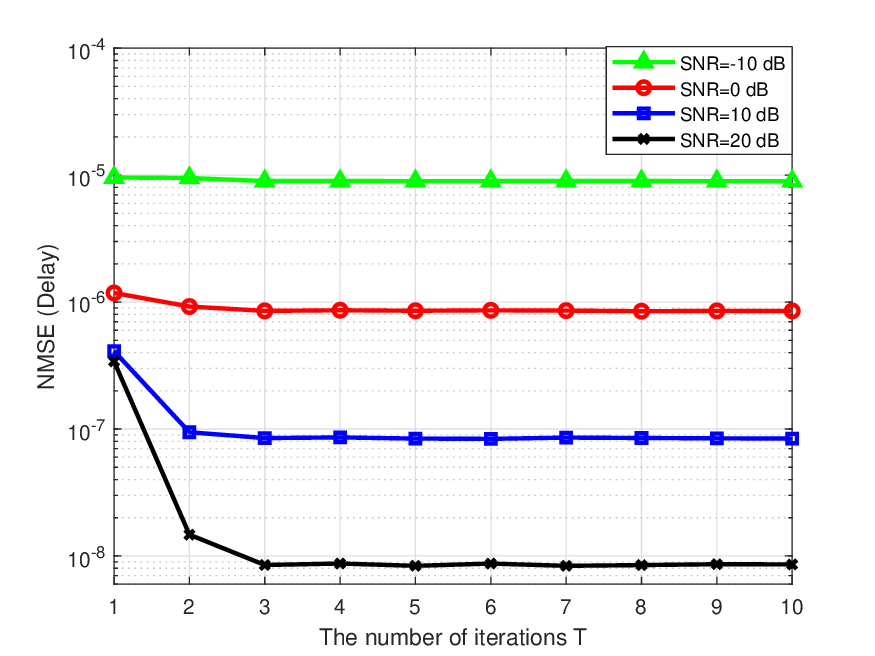}
\captionsetup{justification=raggedright}
\caption{{The NMSE performance of delay estimation versus the number of iterations $T$.}}
\label{T_delay}
\end{figure}
\begin{figure}
\centering
\includegraphics[width=8cm]{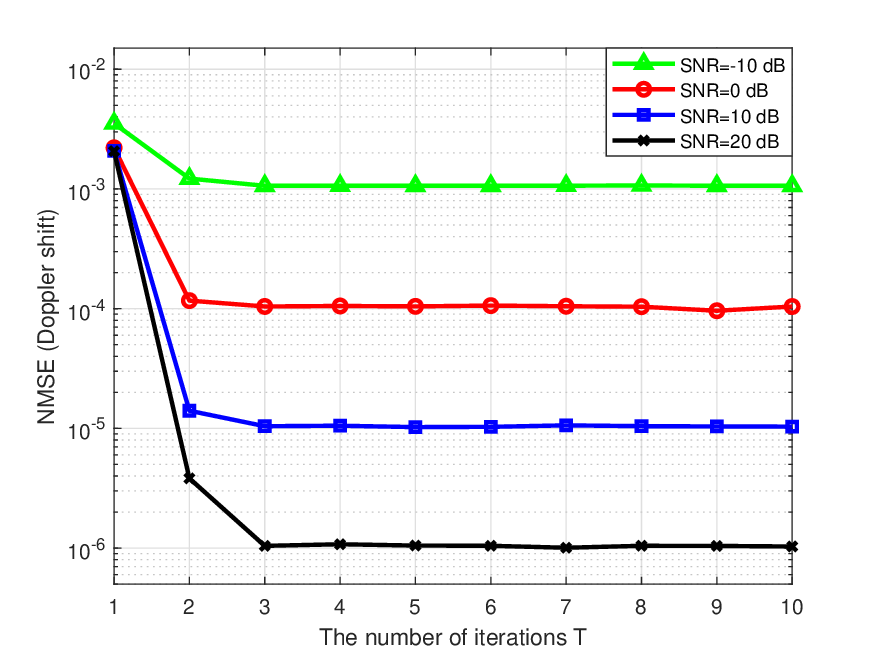}
\captionsetup{justification=raggedright}
\caption{{The NMSE performance of Doppler shift estimation versus the number of iterations $T$.}}
\label{T_doppler}
\end{figure}

We further investigate the AoA estimation performance of our proposed scheme in Fig.~\ref{AOA} for mixed NF and FF targets. Different from AoD estimation, AoA estimation is affected by the presence of near-field targets. We compare the NMSE performance of our proposed scheme with the FF-based scheme, which only considers the plane wavefront, the subspace-based method \cite{W_Zuo}, and the CRLB. We can observe that our proposed scheme achieves superior performance and closely approaches the corresponding CRLB. For the near-field target, the impact of the spherical wavefront propagation becomes significant, which is neglected in the FF-based scheme. Therefore, the FF-based scheme has the worst performance. More importantly, our proposed scheme also has a performance gain compared with the subspace-based method due to the ability to capture the intrinsic multidimensional structure of received signals and the ability to avoid inter-target interference.

\begin{figure}
\centering
\includegraphics[width=8cm]{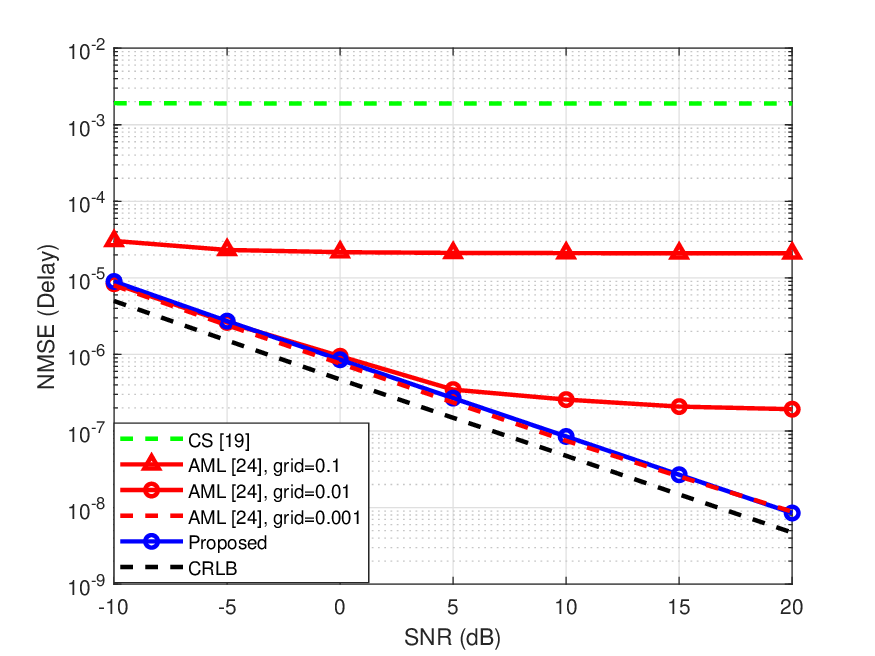}
\caption{{The NMSE performance of delay estimation versus SNR.}}
\label{NMSE_delay}
\end{figure}
\begin{figure}
\centering
\includegraphics[width=8cm]{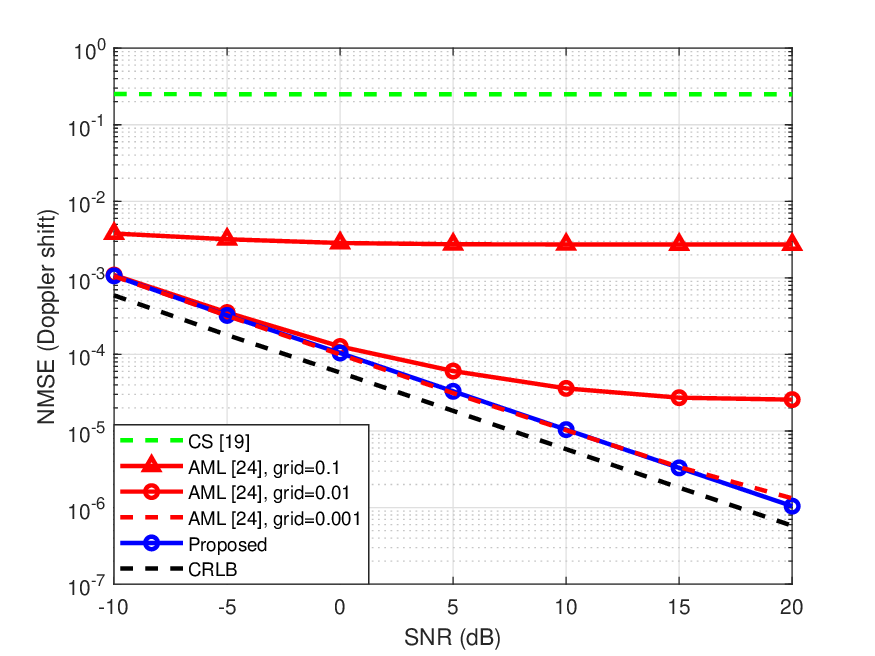}
\caption{{The NMSE performance of Doppler shift estimation versus SNR.}}
\label{NMSE_doppler}
\end{figure}

{Next, we investigate the convergence performance of the proposed Algorithm 2 for delay and Doppler shift estimation. Fig.~\ref{T_delay} and Fig.~\ref{T_doppler} illustrate the NMSE performance of the proposed Algorithm 2 versus the number of iterations $T$ under different SNRs for delay and Doppler shift estimation, respectively. It is obvious that, as the number of iterations increases, the estimation performance of both delay and Doppler shift progressively enhanced and ultimately converged. This implies that our proposed scheme can rapidly converge to a stationary solution. Specifically, our proposed Algorithm 2 only requires around $3$ iterations for convergence. 
In the rest of our experiments, we shall use $T = 3$ unless otherwise noted.}

Fig.~\ref{NMSE_delay} and Fig.~\ref{NMSE_doppler} provide the NMSE performance of the delay and Doppler shift estimation versus SNR, respectively. For comparison, we apply the compressed sensing (CS) method \cite{W_Benzine_CS}, the approximate maximum likelihood (AML) method with different delay and Doppler shift resolutions \cite{ML}, and CRLB as benchmarks. We again notice that the performance of our proposed scheme approaches the associated CRLB. However, the CS method is strictly limited by the sampling rate and bandwidth, which can only estimate the integer components of normalized delay and Doppler shift, resulting in the worst performance. The performance of the AML method is under the control of the searching resolution and has a complexity of $\mathcal{O}\left(R\left( {{l_{\max }} + 1} \right)\left( {2{\alpha _{\max }} + 1} \right)\frac{{N_c^2}}{{{\varepsilon _\tau}{\varepsilon _f}}}\right)$, where ${\varepsilon _\tau}$ is the searching grid resolution of delay and ${\varepsilon _f}$ is the searching grid resolution of Doppler shift. We compare with the performance of AML method \cite{ML} in the case of choosing different resolutions of $\varepsilon _\tau=\varepsilon _f=0.1$, $\varepsilon _\tau=\varepsilon _f=0.01$, and $\varepsilon _\tau=\varepsilon _f=0.001$. It is obvious that higher resolution leads to better estimation performance, but at the cost of increased computational complexity. Our proposed method can achieve the same performance as the case of $\varepsilon _\tau=\varepsilon _f=0.001$ but with much lower complexity. To be specific, the complexity for the case of $\varepsilon _\tau=\varepsilon _f=0.001$ for AML is of order $\mathcal{O}\left(10^{12}\right)$ while our proposed method has the complexity of order $\mathcal{O}\left(10^{4}\right)$ in the same simulation setup.

\section{Conclusion}
In this paper, we proposed a novel angle-delay-Doppler estimation scheme for the AFDM-ISAC system in mixed NF and FF scenarios. Based on the echoes reflected by the targets, the AoDs, AoAs, delays and Doppler shifts can be respectively estimated via a structured tensor decomposition based method. Specifically, we formulated the received ISAC signal as a third-order tensor with factor matrices containing sensing target parameters. We then applied a structured CP decomposition method by employing the Vandermonde nature of the factor matrix and the spatial smoothing. Subsequently, we developed a subspace-based method that avoids high-complexity eigendecomposition for AoA estimation in mixed NF and FF scenario.
 By leveraging the full representation of channel delay-Doppler profile provided by AFDM in the DAF domain, we also proposed a joint delay-Doppler estimation scheme, where the fractional components can be estimated by using one-dimensional iterative process for sensing enhancement with low complexity. Simulation results validated the accuracy of our proposed method and demonstrated that the performance of the proposed scheme outperforms the other benchmark schemes and also close to that of the corresponding CRLB.
 %and outperforms other approaches in terms of accuracy or computational complexity.

{\appendix[Derivation of CRLB]}
The observation tensor $\boldsymbol{\mathcal{Y}} \in {\mathbb{C}^{{G} \times {N_c} \times {K}}}$ has the expression of
\begin{equation}
    \boldsymbol{\mathcal{Y}} = \left[\kern-0.15em\left[ {{\boldsymbol{\gamma }};{{\mathbf{A}}_R},{{\mathbf{B}}_C},{{\mathbf{A}}_T}} 
 \right]\kern-0.15em\right]+\boldsymbol{\mathcal{W}}.\label{Y_crlb}
\end{equation}

Let $\mathbf{A}=\mathbf{A}_R$, $\mathbf{B}=\mathbf{B}_C$ and $\mathbf{C}=\text{diag}\left(\boldsymbol{\gamma }\right)\mathbf{A}_T$. Therefore, the log-likelihood function of $\boldsymbol{\Psi}$ can be written as

\begin{equation}
\begin{split}
  L\left( \boldsymbol{\Psi} \right) &= f\left( {\boldsymbol{\mathcal{Y}};\mathbf{A},\mathbf{B},\mathbf{C}} \right)\\
   &=  - GN_cK\ln \left( {\pi {\sigma ^2}} \right) - \frac{1}{{{\sigma ^2}}}\left\| {{\mathbf{Y}}_{(1)}^T - \left( {{\mathbf{C}} \odot {\mathbf{B}}} \right){{\mathbf{A}}^T}} \right\|_F^2 \\
   &=  - GN_cK\ln \left( {\pi {\sigma ^2}} \right) - \frac{1}{{{\sigma ^2}}}\left\| {{\mathbf{Y}}_{(2)}^T - \left( {{\mathbf{C}} \odot {\mathbf{A}}} \right){{\mathbf{B}}^T}} \right\|_F^2 \\
   &=  - GN_cK\ln \left( {\pi {\sigma ^2}} \right) - \frac{1}{{{\sigma ^2}}}\left\| {{\mathbf{Y}}_{(3)}^T - \left( {{\mathbf{B}} \odot {\mathbf{A}}} \right){{\mathbf{C}}^T}} \right\|_F^2,
\end{split}  
\end{equation}
where $\sigma^2$ is the noise variance. 
The Fisher information matrix (FIM) for $\boldsymbol{\Psi}$ is given by \cite{CRLB_cpd}
\begin{equation}
    \boldsymbol{\Omega} \left( \boldsymbol{\Psi} \right) = \mathbb{E}\left\{ {{{\left( {\frac{{\partial L\left( \boldsymbol{\Psi} \right)}}{{\partial \boldsymbol{\Psi}}}} \right)}^H}\left( {\frac{{\partial L\left( \boldsymbol{\Psi}\right)}}{{\partial \boldsymbol{\Psi}}}} \right)} \right\}.\label{Omega}
\end{equation}
For problems considered in this paper, \eqref{Omega} is equivalent to
\begin{equation}
  \boldsymbol{\Omega}\left( \boldsymbol{\Psi} \right)\hspace{-0.4em}=\hspace{-0.4em}\left[ {\begin{array}{*{20}{c}}
  {{{\boldsymbol{\Omega }}_{{\boldsymbol{\theta \theta }}}}}&{{{\boldsymbol{\Omega }}_{{\boldsymbol{\theta }}{{\mathbf{r}}_{\mathbf{0}}}}}}&{{{\boldsymbol{\Omega }}_{{\boldsymbol{\theta \tau }}}}}&{{{\boldsymbol{\Omega }}_{{\boldsymbol{\theta }}{{\mathbf{f}}_{\mathbf{d}}}}}}&{{{\boldsymbol{\Omega }}_{{\boldsymbol{\theta }}\phi }}}&{{{\boldsymbol{\Omega }}_{{\boldsymbol{\theta }}\boldsymbol{\gamma} }}} \\ 
  {{\boldsymbol{\Omega }}_{{\boldsymbol{\theta }}{{\mathbf{r}}_{\mathbf{0}}}}^H}&{{{\boldsymbol{\Omega }}_{{{\mathbf{r}}_{\mathbf{0}}}{{\mathbf{r}}_{\mathbf{0}}}}}}&{{\boldsymbol{\Omega }}_{{{\mathbf{r}}_{\mathbf{0}}}{\boldsymbol{{\tau }}}}}&{{{\boldsymbol{\Omega }}_{{{\mathbf{r}}_{\mathbf{0}}}{{\mathbf{f}}_{\mathbf{d}}}}}}&{{{\boldsymbol{\Omega }}_{{{\mathbf{r}}_{\mathbf{0}}}\boldsymbol{\phi} }}}&{{{\boldsymbol{\Omega }}_{{{\mathbf{r}}_{\mathbf{0}}}\gamma }}} \\ 
  {{\boldsymbol{\Omega }}_{{\boldsymbol{\theta \tau }}}^H}&{{\boldsymbol{\Omega }}_{{{\mathbf{r}}_{\mathbf{0}}}{\boldsymbol{\tau }}}^H}&{{{\boldsymbol{\Omega }}_{{\boldsymbol{\tau \tau }}}}}&{{{\boldsymbol{\Omega }}_{{\boldsymbol{\tau }}{{\mathbf{f}}_{\mathbf{d}}}}}}&{{{\boldsymbol{\Omega }}_{{\boldsymbol{\tau \phi}} }}}&{{{\boldsymbol{\Omega }}_{{\boldsymbol{\tau \gamma }}}}} \\ 
  {{\boldsymbol{\Omega }}_{{\boldsymbol{\theta }}{{\mathbf{f}}_{\mathbf{d}}}}^H}&{{\boldsymbol{\Omega }}_{{{\mathbf{r}}_{\mathbf{0}}}{{\mathbf{f}}_{\mathbf{d}}}}^H}&{{\boldsymbol{\Omega }}_{{\boldsymbol{\tau }}{{\mathbf{f}}_{\mathbf{d}}}}^H}&{{{\boldsymbol{\Omega }}_{{{\mathbf{f}}_{\mathbf{d}}}{{\mathbf{f}}_{\mathbf{d}}}}}}&{{{\boldsymbol{\Omega }}_{{{\mathbf{f}}_{\mathbf{d}}}\phi }}}&{{{\boldsymbol{\Omega }}_{{{\mathbf{f}}_{\mathbf{d}}}\boldsymbol{\gamma} }}} \\ 
  {{\boldsymbol{\Omega }}_{{\boldsymbol{\theta \phi}} }^H}&{{\boldsymbol{\Omega }}_{{{\mathbf{r}}_{\mathbf{0}}}{\boldsymbol{\phi}}}^H}&{{\boldsymbol{\Omega }}_{{\boldsymbol{\tau \phi}}}^H}&{{\boldsymbol{\Omega }}_{{{\mathbf{f}}_{\mathbf{d}}}\boldsymbol{\phi} }^H}&{{{\boldsymbol{\Omega }}_{\boldsymbol{\phi \phi }}}}&{{{\boldsymbol{\Omega }}_{\boldsymbol{\phi \gamma }}}} \\ 
  {{\boldsymbol{\Omega }}_{{\boldsymbol{\theta \gamma }} }^H}&{{\boldsymbol{\Omega }}_{{{\mathbf{r}}_{\mathbf{0}}}\boldsymbol{\gamma} }^H}&{{\boldsymbol{\Omega }}_{{\boldsymbol{\tau \gamma }}}^H}&{{\boldsymbol{\Omega }}_{{{\mathbf{f}}_{\mathbf{d}}}\boldsymbol{\gamma} }^H}&{{\boldsymbol{\Omega }}^H_{\boldsymbol{\phi \gamma }}}&{{{\boldsymbol{\Omega }}_{\boldsymbol{\gamma \gamma }}}} 
\end{array}} \right].
\end{equation}

We first compute the partial derivative of $ L\left( \boldsymbol{\Psi} \right)$ with respect to $ \boldsymbol{\Psi}$. According to \cite{tensor}, the partial derivative of $L\left( \boldsymbol{\Psi}\right)$ with respect to $\theta_r$ can be expressed as 
\begin{equation}
    \frac{{\partial L\left( \boldsymbol{\Psi} \right)}}{{\partial {\theta _r}}} = {\text{tr}}\left\{ {{{\left( {\frac{{\partial L\left( \boldsymbol{\Psi} \right)}}{{\partial {\mathbf{A}}}}} \right)}^T}\frac{{\partial {\mathbf{A}}}}{{\partial {\theta _r}}} + {{\left( {\frac{{\partial L\left( \boldsymbol{\Psi} \right)}}{{\partial {{\mathbf{A}}^*}}}} \right)}^T}\frac{{\partial {{\mathbf{A}}^*}}}{{\partial {\theta _r}}}} \right\},\label{deri_theta}
\end{equation}
where
\[\frac{{\partial L\left( \boldsymbol{\Psi} \right)}}{{\partial {\mathbf{A}}}} = \frac{1}{{{\sigma ^2}}}{\left( {{\mathbf{Y}}_{(1)}^T - \left( {{\mathbf{C}} \odot {\mathbf{B}}} \right){{\mathbf{A}}^T}} \right)^H}\left( {{\mathbf{C}} \odot {\mathbf{B}}} \right),\]
\[\frac{{\partial L\left( \boldsymbol{\Psi} \right)}}{{\partial {{\mathbf{A}}^*}}} = {\left( {\frac{{\partial L\left( \boldsymbol{\Psi} \right)}}{{\partial {\mathbf{A}}}}} \right)^*},\]
\[\frac{{\partial {\mathbf{A}}}}{{\partial {\theta _r}}} = \left[ {\begin{array}{*{20}{c}}
  {\mathbf{0}}& \cdots &{\mathbf{\tilde a}}_{\theta, r}& \cdots &{\mathbf{0}} 
\end{array}} \right],\]
with ${\mathbf{\tilde a}}_{\theta, r}= \frac{{\partial {{\mathbf{a}}_R}\left( {{\theta _r},{r_{0,r}}} \right)}}{{\partial {\theta _r}}}$. Therefore, \eqref{deri_theta} can be calculated as
\begin{equation}
\frac{{\partial L\left( \boldsymbol{\Psi} \right)}}{{\partial {\theta _r}}} = 2\operatorname{Re} \left( {{\mathbf{e}}_r^T\frac{1}{{{\sigma ^2}}}{{\left( {{\mathbf{C}} \odot {\mathbf{B}}} \right)}^T}{{{\mathbf{\tilde Y}}}_1}^*{\mathbf{\tilde A}_\theta}{{\mathbf{e}}_r}} \right),
\end{equation}
where ${{{\mathbf{\tilde Y}}}_1} =  {{\mathbf{Y}}_{(1)}^T - \left( {{\mathbf{C}} \odot {\mathbf{B}}} \right){{\mathbf{A}}^T}} $, ${{\mathbf{e}}_r}$ is a canonical vector whose non-zero element has the index of $r$, $\operatorname{Re} \left(  \cdot  \right)$ is the real part extraction operator, and ${\mathbf{\tilde A}_\theta} = \left[ {{{{\mathbf{\tilde a}}}_{\theta,1}},{{{\mathbf{\tilde a}}}_{\theta,2}},...,{{{\mathbf{\tilde a}}}_{\theta,R}}} \right]$.

Similarly, the partial derivatives with respect to other parameters can be respectively expressed as
$$\frac{{\partial L\left( \boldsymbol{\Psi} \right)}}{{\partial {r _{0,r}}}} = 2\operatorname{Re} \left( {{\mathbf{e}}_r^T\frac{1}{{{\sigma ^2}}}{{\left( {{\mathbf{C}} \odot {\mathbf{B}}} \right)}^T}{{{\mathbf{\tilde Y}}}_1}^*{\mathbf{\tilde A}_{r _{0}}}{{\mathbf{e}}_r}} \right),
  $$
  $$\frac{{\partial L\left( \boldsymbol{\Psi} \right)}}{{\partial {\tau_r}}} = 2\operatorname{Re} \left( {{\mathbf{e}}_r^T\frac{1}{{{\sigma ^2}}}{{\left( {{\mathbf{C}} \odot {\mathbf{A}}} \right)}^T}{{{\mathbf{\tilde Y}}}_2}^*{{{\mathbf{\tilde B}}}_{\tau}}{{\mathbf{e}}_r}} \right) \hfill,$$
\[\begin{gathered}
% \frac{{\partial L\left( \boldsymbol{\Psi} \right)}}{{\partial {r _{0,r}}}} = 2\operatorname{Re} \left( {{\mathbf{e}}_r^T\frac{1}{{{\sigma ^2}}}{{\left( {{\mathbf{C}} \odot {\mathbf{B}}} \right)}^T}{{{\mathbf{\tilde Y}}}_1}^*{\mathbf{\tilde A}_{r _{0}}}{{\mathbf{e}}_r}} \right)\hfill, \\
%   \frac{{\partial L\left( \boldsymbol{\Psi} \right)}}{{\partial {\tau_r}}} = 2\operatorname{Re} \left( {{\mathbf{e}}_r^T\frac{1}{{{\sigma ^2}}}{{\left( {{\mathbf{C}} \odot {\mathbf{A}}} \right)}^T}{{{\mathbf{\tilde Y}}}_2}^*{{{\mathbf{\tilde B}}}_{\tau}}{{\mathbf{e}}_r}} \right) \hfill, \\
  \frac{{\partial L\left( \boldsymbol{\Psi} \right)}}{{\partial {f _{d,r}}}} = 2\operatorname{Re} \left( {{\mathbf{e}}_r^T\frac{1}{{{\sigma ^2}}}{{\left( {{\mathbf{C}} \odot {\mathbf{A}}} \right)}^T}{{{\mathbf{\tilde Y}}}_2}^*{{{\mathbf{\tilde B}}}_{f} }{{\mathbf{e}}_r}} \right), \hfill \\
  \frac{{\partial L\left( \boldsymbol{\Psi} \right)}}{{\partial {\phi_r}}} = 2\operatorname{Re} \left( {{\mathbf{e}}_r^T\frac{1}{{{\sigma ^2}}}{{\left( {{\mathbf{B}} \odot {\mathbf{A}}} \right)}^T}{{{\mathbf{\tilde Y}}}_3}^*{\mathbf{\tilde C}}{{\mathbf{e}}_r}} \right), \hfill \\
  \frac{{\partial L\left( \boldsymbol{\Psi} \right)}}{{\partial {\gamma _r}}} = {\mathbf{e}}_r^T\frac{1}{{{\sigma ^2}}}{\left( {{\mathbf{B}} \odot {\mathbf{A}}} \right)^T}{{{\mathbf{\tilde Y}}}_3}^*{{\mathbf{A}}_T}{{\mathbf{e}}_r}, \hfill \\ 
\end{gathered} \]
with ${{\mathbf{\tilde Y}}}_2 = {\mathbf{Y}}_{(2)}^T - \left( {{\mathbf{C}} \odot {\mathbf{A}}} \right){{\mathbf{B}}^T}$, ${{{\mathbf{\tilde Y}}}_3} = {\mathbf{Y}}_{(3)}^T - \left( {{\mathbf{B}} \odot {\mathbf{A}}} \right){{\mathbf{C}}^T}$ and ${{{\mathbf{\tilde A}}}_{r_0}} = \left[ {{{{\mathbf{\tilde a}}}_{r_0,1}},{{{\mathbf{\tilde a}}}_{r_0,2}},...,{{{\mathbf{\tilde a}}}_{r_0,R}}} \right]$, ${{{\mathbf{\tilde B}}}_{\tau}} = \left[ {{{{\mathbf{\tilde b}}}_{\tau,1}},{{{\mathbf{\tilde b}}}_{\tau,2}},...,{{{\mathbf{\tilde b}}}_{\tau,R}}} \right]$, ${{{\mathbf{\tilde B}}}_f } = \left[ {{{{\mathbf{\tilde b}}}_{f ,1}},{{{\mathbf{\tilde b}}}_{f ,2}},...,{{{\mathbf{\tilde b}}}_{f ,R}}} \right]$, ${\mathbf{\tilde C}} = \left[ {{{{\mathbf{\tilde c}}}_1},{{{\mathbf{\tilde c}}}_2},...,{{{\mathbf{\tilde c}}}_R}} \right]$, where ${\mathbf{\tilde a}}_{r_0, r}= \frac{{\partial {{\mathbf{a}}_R}\left( {{\theta _r},{r_{0,r}}} \right)}}{{\partial{r_{0,r}}}}$, ${{\mathbf{\tilde b}}}_{\tau,r}=\frac{{\partial {{\mathbf{b}}}\left( {{\beta _r},\nu_r} \right)}}{{\partial{\tau_r}}}$, ${{\mathbf{\tilde b}}}_{f,r}=\frac{{\partial {{\mathbf{b}}}\left( {{\beta _r},\nu_r} \right)}}{{\partial{f_{d,r}}}}$ and ${{{\mathbf{\tilde c}}}_r}=\frac{\gamma_r{\partial {{\mathbf{a}}_T}\left( \phi_r \right)}}{{\partial{\phi_r}}}$ for $r=1,...,R$. %${{{\mathbf{\tilde c}}}_r}=-j\pi \cos \left( {{\phi _r}} \right){\text{diag}}\left( {0,1,...,K - 1} \right){\mathbf{a}}\left( {{\phi _r}} \right)$.

Next, we do the calculation for components in the principal minors of the FIM. The $\left(r_1,r_2\right)$-th entry of
\begin{equation}
    {{{\boldsymbol{\Omega }}_{{\boldsymbol{\theta \theta }}}}}=\mathbb{E}\left\{ {{{\left( {\frac{{\partial L\left( \boldsymbol{\Psi} \right)}}{{\partial \boldsymbol{\theta}}}} \right)}^H}\left( {\frac{{\partial L\left( \boldsymbol{\Psi} \right)}}{{\partial \boldsymbol{\theta}}}} \right)} \right\}
\end{equation}
is given by 
\begin{equation}
    \mathbb{E}\left\{ {{{\left( {\frac{{\partial L\left( \boldsymbol{\Psi} \right)}}{{\partial {\theta _{{r_1}}}}}} \right)}^*}\left( {\frac{{\partial L\left( \boldsymbol{\Psi}\right)}}{{\partial {\theta _{{r_2}}}}}} \right)} \right\} =  {{\mathbf{C}}_{{{\mathbf{n}}_{a_\theta}}}\left( {m,n} \right)}+{{\mathbf{C}}^*_{{{\mathbf{n}}_{a_\theta}}}\left( {m,n} \right)},
\end{equation}
where $m=R(r_1-1)+r_1$, $n=R(r_2-1)+r_2$ and ${{\mathbf{C}}_{{{\mathbf{n}}_{a_\theta}}}} = \mathbb{E}\left\{ {{{\mathbf{n}}_{a_\theta}}{{\mathbf{n}}^H_{a_\theta}}} \right\}$ with ${{\mathbf{n}}_{a_\theta}} = \frac{1}{\sigma^2}\left( {{{\mathbf{\tilde A}_\theta}^T} \otimes {{\left( {{\mathbf{C}} \odot {\mathbf{B}}} \right)}^T}} \right)\text{vec}\left( {{\mathbf{W}}_{\left( 1 \right)}^H} \right)$.

Similarly, other components in the principal minors of ${\boldsymbol{\Omega }}\left( \boldsymbol{\Psi} \right)$ can also be calculated by
$$\mathbb{E}\left\{ {{{\left( {\frac{{\partial L\left( \boldsymbol{\Psi} \right)}}{{\partial {r_{0,{r_1}}}}}} \right)}^*}\left( {\frac{{\partial L\left( \boldsymbol{\Psi} \right)}}{{\partial {r_{0,{r_2}}}}}} \right)} \right\}={{\mathbf{C}}_{{{\mathbf{n}}_{{a_{{r_{_0}}}}}}}}\hspace{-0.4em}\left( {m,n} \right) + {{\mathbf{C}}_{{{\mathbf{n}}_{{a_{{r_{_0}}}}}}}}\hspace{-0.4em}{\left( {m,n} \right)^*},$$
$$\mathbb{E}\left\{ {{{\left( {\frac{{\partial L\left( \boldsymbol{\Psi} \right)}}{{\partial {\tau _{{r_1}}}}}} \right)}^*}\left( {\frac{{\partial L\left( \boldsymbol{\Psi} \right)}}{{\partial {\tau _{{r_2}}}}}} \right)} \right\} = {{\mathbf{C}}_{{{\mathbf{n}}_{{b_\tau }}}}}\left( {m,n} \right) + {{\mathbf{C}}_{{{\mathbf{n}}_{{b_\tau }}}}}{\left( {m,n} \right)^*},$$
$$\mathbb{E}\left\{ {{{\left( {\frac{{\partial L\left( \boldsymbol{\Psi} \right)}}{{\partial {f_{d,{r_1}}}}}} \right)}^*}\left( {\frac{{\partial L\left( \boldsymbol{\Psi} \right)}}{{\partial {f_{d,{r_2}}}}}} \right)} \right\} = {{\mathbf{C}}_{{{\mathbf{n}}_{{b_f}}}}}\left( {m,n} \right) + {{\mathbf{C}}_{{{\mathbf{n}}_{{b_f}}}}}{\left( {m,n} \right)^*},$$
\[\begin{gathered}
  % \mathbb{E}\left\{ {{{\left( {\frac{{\partial L\left( \boldsymbol{\Psi} \right)}}{{\partial {r_{0,{r_1}}}}}} \right)}^*}\left( {\frac{{\partial L\left( \boldsymbol{\Psi} \right)}}{{\partial {r_{0,{r_2}}}}}} \right)} \right\}={{\mathbf{C}}_{{{\mathbf{n}}_{{a_{{r_{_0}}}}}}}}\hspace{-0.4em}\left( {m,n} \right) + {{\mathbf{C}}_{{{\mathbf{n}}_{{a_{{r_{_0}}}}}}}}\hspace{-0.4em}{\left( {m,n} \right)^*}, \hfill \\
  % \mathbb{E}\left\{ {{{\left( {\frac{{\partial L\left( \boldsymbol{\Psi} \right)}}{{\partial {\tau _{{r_1}}}}}} \right)}^*}\left( {\frac{{\partial L\left( \boldsymbol{\Psi} \right)}}{{\partial {\tau _{{r_2}}}}}} \right)} \right\} = {{\mathbf{C}}_{{{\mathbf{n}}_{{b_\tau }}}}}\left( {m,n} \right) + {{\mathbf{C}}_{{{\mathbf{n}}_{{b_\tau }}}}}{\left( {m,n} \right)^*}, \hfill \\
  % \mathbb{E}\left\{ {{{\left( {\frac{{\partial L\left( \boldsymbol{\Psi} \right)}}{{\partial {f_{d,{r_1}}}}}} \right)}^*}\left( {\frac{{\partial L\left( \boldsymbol{\Psi} \right)}}{{\partial {f_{d,{r_2}}}}}} \right)} \right\} = {{\mathbf{C}}_{{{\mathbf{n}}_{{b_f}}}}}\left( {m,n} \right) + {{\mathbf{C}}_{{{\mathbf{n}}_{{b_f}}}}}{\left( {m,n} \right)^*}, hfill \\
  \mathbb{E}\left\{ {{{\left( {\frac{{\partial L\left( \boldsymbol{\Psi} \right)}}{{\partial {\phi _{{r_1}}}}}} \right)}^*}\left( {\frac{{\partial L\left( \boldsymbol{\Psi} \right)}}{{\partial {\phi _{{r_2}}}}}} \right)} \right\} = {{\mathbf{C}}_{{{\mathbf{n}}_\phi }}}\left( {m,n} \right) + {{\mathbf{C}}_{{{\mathbf{n}}_\phi }}}{\left( {m,n} \right)^*}, \hfill \\ 
  \mathbb{E}\left\{ {{{\left( {\frac{{\partial L\left( \boldsymbol{\Psi} \right)}}{{\partial {\gamma _{{r_1}}}}}} \right)}^*}\left( {\frac{{\partial L\left( \boldsymbol{\Psi} \right)}}{{\partial {\gamma _{{r_2}}}}}} \right)} \right\} ={{\mathbf{C}}_{{{\mathbf{n}}_\gamma }}}{\left( {m,n} \right)^*}. \hfill \\ 
\end{gathered} \]

Similarly to $\mathbf{n}_{a_\theta}$,
$${{\mathbf{n}}_{a_{r_0}}} = \frac{1}{\sigma^2}\left( {{{\mathbf{\tilde A}_{r_0}}^T} \otimes {{\left( {{\mathbf{C}} \odot {\mathbf{B}}} \right)}^T}} \right)\text{vec}\left( {{\mathbf{W}}_{\left( 1 \right)}^H} \right),$$
$${{\mathbf{n}}_{b_\tau}} = \left( {{{{\mathbf{\tilde B}_\tau}}^T} \otimes {{\left( {{\mathbf{C}} \odot {\mathbf{A}}} \right)}^T}} \right)\text{vec}\left( {{\mathbf{W}}_{\left( 2 \right)}^H} \right),$$
$${{\mathbf{n}}_{b_f}} = \left( {{{{\mathbf{\tilde B}_f}}^T} \otimes {{\left( {{\mathbf{C}} \odot {\mathbf{A}}} \right)}^T}} \right)\text{vec}\left( {{\mathbf{W}}_{\left( 2 \right)}^H} \right),$$
$${{\mathbf{n}}_{\phi}} = \left( {{{{\mathbf{\tilde C}}}^T} \otimes {{\left( {{\mathbf{B}} \odot {\mathbf{A}}} \right)}^T}} \right)\text{vec}\left( {{\mathbf{W}}_{\left( 3 \right)}^H} \right),$$
$${{\mathbf{n}}_{\gamma}} = \left( {{{{\mathbf{A}}}^T_T} \otimes {{\left( {{\mathbf{B}} \odot {\mathbf{A}}} \right)}^T}} \right)\text{vec}\left( {{\mathbf{W}}_{\left( 3 \right)}^H} \right).$$

For the elements in the off-principal minors, we can obtain 
$$\mathbb{E}\left\{ {{{\left( {\frac{{\partial L\left( \boldsymbol{\Psi} \right)}}{{\partial {\theta _{{r_1}}}}}} \right)}^*}\left( {\frac{{\partial L\left( \boldsymbol{\Psi} \right)}}{{\partial {r_{0,{r_2}}}}}} \right)} \right\} = 2\operatorname{Re} \left( {{{\mathbf{C}}_{{{\mathbf{n}}_{{a_{_\theta }}}},{{\mathbf{n}}_{{a_{{r_{_0}}}}}}}}\left( {m,n} \right)} \right), $$
$$\mathbb{E}\left\{ {{{\left( {\frac{{\partial L\left( \boldsymbol{\Psi} \right)}}{{\partial {\theta _{{r_1}}}}}} \right)}^*}\left( {\frac{{\partial L\left( \boldsymbol{\Psi}\right)}}{{\partial {\tau _{{r_2}}}}}} \right)} \right\} = 2\operatorname{Re} \left( {{{\mathbf{C}}_{{{\mathbf{n}}_{{a_{_\theta }}}},{{\mathbf{n}}_{{b_\tau }}}}}\left( {m,n} \right)} \right),$$
$$ \mathbb{E}\left\{ {{{\left( {\frac{{\partial L\left( \boldsymbol{\Psi} \right)}}{{\partial {\theta _{{r_1}}}}}} \right)}^*}\left( {\frac{{\partial L\left( \boldsymbol{\Psi} \right)}}{{\partial {f_{{d,r_2}}}}}} \right)} \right\} = 2\operatorname{Re} \left( {{{\mathbf{C}}_{{{\mathbf{n}}_{{a_{_\theta }}}},{{\mathbf{n}}_{{b_f}}}}}\left( {m,n} \right)} \right),$$
$$ \mathbb{E}\left\{ {{{\left( {\frac{{\partial L\left( \boldsymbol{\Psi} \right)}}{{\partial {\theta _{{r_1}}}}}} \right)}^*}\left( {\frac{{\partial L\left( \boldsymbol{\Psi} \right)}}{{\partial {\phi_{{r_2}}}}}} \right)} \right\} = 2\operatorname{Re} \left( {{{\mathbf{C}}_{{{\mathbf{n}}_{{a_{_\theta }}}},{{\mathbf{n}}_{\phi}}}}\left( {m,n} \right)} \right),$$
$$\mathbb{E}\left\{ {{{\left( {\frac{{\partial L\left( \boldsymbol{\Psi} \right)}}{{\partial {\theta _{{r_1}}}}}} \right)}^*}\left( {\frac{{\partial L\left( \boldsymbol{\Psi} \right)}}{{\partial {\gamma _{{r_2}}}}}} \right)} \right\} = {{\mathbf{C}}_{{{\mathbf{n}}_{{a_{_\theta }}}},{{\mathbf{n}}_\gamma }}}{\left( {m,n} \right)^*},$$
$$\mathbb{E}\left\{ {{{\left( {\frac{{\partial L\left( \boldsymbol{\Psi} \right)}}{{\partial {r_{0,{r_1}}}}}} \right)}^*}\left( {\frac{{\partial L\left( \boldsymbol{\Psi} \right)}}{{\partial {\tau _{{r_2}}}}}} \right)} \right\} = 2\operatorname{Re} \left( {{{\mathbf{C}}_{{{\mathbf{n}}_{{a_{{r_{_0}}}}}},{{\mathbf{n}}_{{b_\tau }}}}}\left( {m,n} \right)} \right),$$
$$\mathbb{E}\left\{ {{{\left( {\frac{{\partial L\left( \boldsymbol{\Psi} \right)}}{{\partial {r_{0,{r_1}}}}}} \right)}^*}\left( {\frac{{\partial L\left( \boldsymbol{\Psi} \right)}}{{\partial {f_{{d,r_2}}}}}} \right)} \right\} = 2\operatorname{Re} \left( {{{\mathbf{C}}_{{{\mathbf{n}}_{{a_{{r_{_0}}}}}},{{\mathbf{n}}_{{b_f}}}}}\left( {m,n} \right)} \right),$$
$$\mathbb{E}\left\{ {{{\left( {\frac{{\partial L\left( \boldsymbol{\Psi} \right)}}{{\partial {r_{0,{r_1}}}}}} \right)}^*}\left( {\frac{{\partial L\left( \boldsymbol{\Psi} \right)}}{{\partial {\phi _{{r_2}}}}}} \right)} \right\} = 2\operatorname{Re} \left( {{{\mathbf{C}}_{{{\mathbf{n}}_{{a_{{r_{_0}}}}}},{{\mathbf{n}}_{{\phi }}}}}\left( {m,n} \right)} \right),$$
$$\mathbb{E}\left\{ {{{\left( {\frac{{\partial L\left( \boldsymbol{\Psi} \right)}}{{\partial {r_{0,{r_1}}}}}} \right)}^*}\left( {\frac{{\partial L\left( \boldsymbol{\Psi} \right)}}{{\partial {\gamma _{{r_2}}}}}} \right)} \right\} = {{\mathbf{C}}_{{{\mathbf{n}}_{{a_{{r_{_0}}}}}},{{\mathbf{n}}_\gamma }}}{\left( {m,n} \right)^*},$$
$$\mathbb{E}\left\{ {{{\left( {\frac{{\partial L\left( \boldsymbol{\Psi} \right)}}{{\partial {\tau _{{r_1}}}}}} \right)}^*}\left( {\frac{{\partial L\left( \boldsymbol{\Psi} \right)}}{{\partial {f_{{d,r_2}}}}}} \right)} \right\} = 2\operatorname{Re} \left( {{{\mathbf{C}}_{{{\mathbf{n}}_{{b_\tau }}},{{\mathbf{n}}_{{b_f}}}}}\left( {m,n} \right)} \right),$$
$$\mathbb{E}\left\{ {{{\left( {\frac{{\partial L\left( \boldsymbol{\Psi} \right)}}{{\partial {\tau _{{r_1}}}}}} \right)}^*}\left( {\frac{{\partial L\left( \boldsymbol{\Psi} \right)}}{{\partial {\phi_{{r_2}}}}}} \right)} \right\} = 2\operatorname{Re} \left( {{{\mathbf{C}}_{{{\mathbf{n}}_{{b_\tau }}},{{\mathbf{n}}_{{\phi}}}}}\left( {m,n} \right)} \right),$$
$$\mathbb{E}\left\{ {{{\left( {\frac{{\partial L\left( \boldsymbol{\Psi} \right)}}{{\partial {\tau _{{r_1}}}}}} \right)}^*}\left( {\frac{{\partial L\left( \boldsymbol{\Psi} \right)}}{{\partial {\gamma _{{r_2}}}}}} \right)} \right\} = {{\mathbf{C}}_{{{\mathbf{n}}_{{b_\tau }}},{{\mathbf{n}}_\gamma }}}{\left( {m,n} \right)^*},$$
$$\mathbb{E}\left\{ {{{\left( {\frac{{\partial L\left( \boldsymbol{\Psi} \right)}}{{\partial {f _{{d,r_1}}}}}} \right)}^*}\left( {\frac{{\partial L\left( \boldsymbol{\Psi} \right)}}{{\partial {\phi_{{r_2}}}}}} \right)} \right\} = 2\operatorname{Re} \left( {{{\mathbf{C}}_{{{\mathbf{n}}_{{b_f }}},{{\mathbf{n}}_{{\phi}}}}}\left( {m,n} \right)} \right),$$
$$\mathbb{E}\left\{ {{{\left( {\frac{{\partial L\left( \boldsymbol{\Psi} \right)}}{{\partial {f_{{d,r_1}}}}}} \right)}^*}\left( {\frac{{\partial L\left( \boldsymbol{\Psi} \right)}}{{\partial {\gamma _{{r_2}}}}}} \right)} \right\} = {{\mathbf{C}}_{{{\mathbf{n}}_{{b_f}}},{{\mathbf{n}}_\gamma }}}{\left( {m,n} \right)^*},$$
$$\mathbb{E}\left\{ {{{\left( {\frac{{\partial L\left( \boldsymbol{\Psi} \right)}}{{\partial {\phi _{{r_1}}}}}} \right)}^*}\left( {\frac{{\partial L\left( \boldsymbol{\Psi} \right)}}{{\partial {\gamma _{{r_2}}}}}} \right)} \right\} = {{\mathbf{C}}_{{{\mathbf{n}}_{{\phi }}},{{\mathbf{n}}_\gamma }}}{\left( {m,n} \right)^*},$$
where 
\[\begin{gathered}
  {{\mathbf{C}}_{{{\mathbf{n}}_{{a_{_\theta }}}},{{\mathbf{n}}_{{a_{{r_{0}}}}}}}}\hspace{-0.4em}=\! \frac{1}{{{\sigma ^2}}}{\left( {{{{\mathbf{\tilde A}}}_\theta } \otimes \left( {{\mathbf{C}} \odot {\mathbf{B}}} \right)} \right)^T}\hspace{-0.3em}\left( {{\mathbf{\tilde A}}_{{r_0}}^ *  \otimes {{\left( {{\mathbf{C}} \odot {\mathbf{B}}} \right)}^*}} \right), \hfill \\
  {{\mathbf{C}}_{{{\mathbf{n}}_{{a_{_\theta }}}},{{\mathbf{n}}_{{b_\tau }}}}}\hspace{-0.4em}=\!\frac{1}{{{\sigma ^4}}}{\left( {{{{\mathbf{\tilde A}}}_\theta } \otimes \left( {{\mathbf{C}} \odot {\mathbf{B}}} \right)} \right)^T}\hspace{-0.3em}{{\mathbf{C}}_{{{\mathbf{w}}_{\mathbf{1}}}{\mathbf{,}}{{\mathbf{w}}_{\mathbf{2}}}}}\hspace{-0.3em}\left( {{\mathbf{\tilde B}}_\tau ^ *  \otimes {{\left( {{\mathbf{C}} \odot {\mathbf{A}}} \right)}^*}} \right), \hfill \\
  {{\mathbf{C}}_{{{\mathbf{n}}_{{a_{_\theta }}}},{{\mathbf{n}}_{{b_f}}}}}\hspace{-0.4em}=\!\frac{1}{{{\sigma ^4}}}{\left( {{{{\mathbf{\tilde A}}}_\theta } \otimes \left( {{\mathbf{C}} \odot {\mathbf{B}}} \right)} \right)^T}\hspace{-0.3em}{{\mathbf{C}}_{{{\mathbf{w}}_{\mathbf{1}}}{\mathbf{,}}{{\mathbf{w}}_{\mathbf{2}}}}}\hspace{-0.3em}\left( {{\mathbf{\tilde B}}_f^ *  \otimes {{\left( {{\mathbf{C}} \odot {\mathbf{A}}} \right)}^*}} \right), \hfill \\
  {{\mathbf{C}}_{{{\mathbf{n}}_{{a_{_\theta }}}},{{\mathbf{n}}_{{\phi}}}}}\hspace{-0.4em}=\!\frac{1}{{{\sigma ^4}}}{\left( {{{{\mathbf{\tilde A}}}_\theta } \otimes \left( {{\mathbf{C}} \odot {\mathbf{B}}} \right)} \right)^T}\hspace{-0.3em}{{\mathbf{C}}_{{{\mathbf{w}}_{\mathbf{1}}}{\mathbf{,}}{{\mathbf{w}}_{\mathbf{3}}}}}\hspace{-0.3em}\left( {{\mathbf{\tilde C}}^ *  \otimes {{\left( {{\mathbf{B}} \odot {\mathbf{A}}} \right)}^*}} \right), \hfill \\
  {{\mathbf{C}}_{{{\mathbf{n}}_{{a_{_\theta }}}},{{\mathbf{n}}_\gamma }}}\hspace{-0.4em}=\!\frac{1}{{{\sigma ^4}}}{\left( {{{{\mathbf{\tilde A}}}_\theta } \otimes \left( {{\mathbf{C}} \odot {\mathbf{B}}} \right)} \right)^T}\hspace{-0.3em}{{\mathbf{C}}_{{{\mathbf{w}}_{\mathbf{1}}}{\mathbf{,}}{{\mathbf{w}}_3}}}\hspace{-0.3em}\left( {{\mathbf{A}}_T^ *  \otimes {{\left( {{\mathbf{B}} \odot {\mathbf{A}}} \right)}^*}} \right), \hfill \\
  {{\mathbf{C}}_{{{\mathbf{n}}_{{a_{{r_{0}}}}}},{{\mathbf{n}}_{{b_\tau }}}}}\hspace{-0.4em}=\!\frac{1}{{{\sigma ^4}}}{\left( {{{{\mathbf{\tilde A}}}_{{r_0}}} \otimes \left( {{\mathbf{C}} \odot {\mathbf{B}}} \right)} \right)^T}\hspace{-0.3em}{{\mathbf{C}}_{{{\mathbf{w}}_{\mathbf{1}}}{\mathbf{,}}{{\mathbf{w}}_{\mathbf{2}}}}}\hspace{-0.3em}\left( {{\mathbf{\tilde B}}_\tau ^ *  \otimes {{\left( {{\mathbf{C}} \odot {\mathbf{A}}} \right)}^*}} \right), \hfill \\
  {{\mathbf{C}}_{{{\mathbf{n}}_{{a_{{r_{0}}}}}},{{\mathbf{n}}_{{b_f}}}}}\hspace{-0.4em}=\!\frac{1}{{{\sigma ^4}}}{\left( {{{{\mathbf{\tilde A}}}_{{r_0}}} \otimes \left( {{\mathbf{C}} \odot {\mathbf{B}}} \right)} \right)^T}\hspace{-0.3em}{{\mathbf{C}}_{{{\mathbf{w}}_{\mathbf{1}}}{\mathbf{,}}{{\mathbf{w}}_{\mathbf{2}}}}}\hspace{-0.3em}\left( {{\mathbf{\tilde B}}_f^ *  \otimes {{\left( {{\mathbf{C}} \odot {\mathbf{A}}} \right)}^*}} \right), \hfill \\
  {{\mathbf{C}}_{{{\mathbf{n}}_{{a_{{r_{0}}}}}},{{\mathbf{n}}_{{\phi}}}}}\hspace{-0.4em}=\!\frac{1}{{{\sigma ^4}}}{\left( {{{{\mathbf{\tilde A}}}_{{r_0}}} \otimes \left( {{\mathbf{C}} \odot {\mathbf{B}}} \right)} \right)^T}\hspace{-0.3em}{{\mathbf{C}}_{{{\mathbf{w}}_{\mathbf{1}}}{\mathbf{,}}{{\mathbf{w}}_{\mathbf{3}}}}}\hspace{-0.3em}\left( {{\mathbf{\tilde C}}^ *  \otimes {{\left( {{\mathbf{B}} \odot {\mathbf{A}}} \right)}^*}} \right), \hfill \\
  {{\mathbf{C}}_{{{\mathbf{n}}_{{a_{{r_{0}}}}}},{{\mathbf{n}}_\gamma }}}\hspace{-0.4em}=\!\frac{1}{{{\sigma ^4}}}{\left( {{{{\mathbf{\tilde A}}}_{{r_0}}} \otimes \left( {{\mathbf{C}} \odot {\mathbf{B}}} \right)} \right)^T}\hspace{-0.3em}{{\mathbf{C}}_{{{\mathbf{w}}_{\mathbf{1}}}{\mathbf{,}}{{\mathbf{w}}_3}}}\hspace{-0.3em}\left( {{\mathbf{A}}_T^ *  \otimes {{\left( {{\mathbf{B}} \odot {\mathbf{A}}} \right)}^*}} \right), \hfill \\
  {{\mathbf{C}}_{{{\mathbf{n}}_{{b_\tau }}},{{\mathbf{n}}_{{b_f}}}}}\hspace{-0.4em}=\!\frac{1}{{{\sigma ^2}}}{\left( {{{{\mathbf{\tilde B}}}_\tau } \otimes \left( {{\mathbf{C}} \odot {\mathbf{A}}} \right)} \right)^T}\hspace{-0.3em}\left( {{\mathbf{\tilde B}}_f^ *  \otimes {{\left( {{\mathbf{C}} \odot {\mathbf{A}}} \right)}^*}} \right), \hfill \\
  {{\mathbf{C}}_{{{\mathbf{n}}_{{b_\tau }}},{{\mathbf{n}}_\phi }}}\hspace{-0.4em}=\!\frac{1}{{{\sigma ^4}}}{\left( {{{{\mathbf{\tilde B}}}_\tau } \otimes \left( {{\mathbf{C}} \odot {\mathbf{A}}} \right)} \right)^T}\hspace{-0.3em}{{\mathbf{C}}_{{{\mathbf{w}}_2}{\mathbf{,}}{{\mathbf{w}}_3}}}\hspace{-0.3em}\left( {\tilde{\mathbf{C}}^ *  \otimes {{\left( {{\mathbf{B}} \odot {\mathbf{A}}} \right)}^*}} \right), \hfill \\
  {{\mathbf{C}}_{{{\mathbf{n}}_{{b_\tau }}},{{\mathbf{n}}_\gamma }}}\hspace{-0.4em}=\!\frac{1}{{{\sigma ^4}}}{\left( {{{{\mathbf{\tilde B}}}_\tau } \otimes \left( {{\mathbf{C}} \odot {\mathbf{A}}} \right)} \right)^T}\hspace{-0.3em}{{\mathbf{C}}_{{{\mathbf{w}}_2}{\mathbf{,}}{{\mathbf{w}}_3}}}\hspace{-0.3em}\left( {{\mathbf{A}}_T^ *  \otimes {{\left( {{\mathbf{B}} \odot {\mathbf{A}}} \right)}^*}} \right), \hfill \\
      {{\mathbf{C}}_{{{\mathbf{n}}_{{b_f}}},{{\mathbf{n}}_\phi }}}\hspace{-0.4em}=\!\frac{1}{{{\sigma ^4}}}{\left( {{{{\mathbf{\tilde B}}}_f} \otimes \left( {{\mathbf{C}} \odot {\mathbf{A}}} \right)} \right)^T}\hspace{-0.3em}{{\mathbf{C}}_{{{\mathbf{w}}_2}{\mathbf{,}}{{\mathbf{w}}_3}}}\hspace{-0.3em}\left( {\tilde{\mathbf{C}}^ *  \otimes {{\left( {{\mathbf{B}} \odot {\mathbf{A}}} \right)}^*}} \right), \hfill \\ 
  {{\mathbf{C}}_{{{\mathbf{n}}_{{b_f}}},{{\mathbf{n}}_\gamma }}}\hspace{-0.4em}=\!\frac{1}{{{\sigma ^4}}}{\left( {{{{\mathbf{\tilde B}}}_f} \otimes \left( {{\mathbf{C}} \odot {\mathbf{A}}} \right)} \right)^T}\hspace{-0.3em}{{\mathbf{C}}_{{{\mathbf{w}}_2}{\mathbf{,}}{{\mathbf{w}}_3}}}\hspace{-0.3em}\left( {{\mathbf{A}}_T^ *  \otimes {{\left( {{\mathbf{B}} \odot {\mathbf{A}}} \right)}^*}} \right), \hfill \\ 
  {{\mathbf{C}}_{{{\mathbf{n}}_{{\phi}}},{{\mathbf{n}}_\gamma }}}\hspace{-0.4em}=\!\frac{1}{{{\sigma ^2}}}{\left( {{{{\mathbf{\tilde C}}}} \otimes \left( {{\mathbf{B}} \odot {\mathbf{A}}} \right)} \right)^T}\hspace{-0.3em}\left( {{\mathbf{A}}_T^ *  \otimes {{\left( {{\mathbf{B}} \odot {\mathbf{A}}} \right)}^*}} \right), \hfill \\ 
\end{gathered} \]
in which the covariance matrix for the noise is expressed as

\begin{equation}
    {{\mathbf{C}}_{{{\mathbf{w}}_p},{{\mathbf{w}}_q}}} = \mathbb{E}\left\{ {\text{vec}\left( {{\mathbf{W}}_{\left( p \right)}^H} \right)\text{vec}{{\left( {{\mathbf{W}}_{\left( q \right)}^T} \right)}^T}} \right\},p \ne q.
\end{equation}

Finally, the CRLB for the estimated parameters can be calculated by \eqref{CRLB}.

\bibliographystyle{IEEEtran}
%\small
\footnotesize
\bibliography{ref_AFDM}

\begin{thebibliography}{10}
\providecommand{\url}[1]{#1}
\csname url@samestyle\endcsname
\providecommand{\newblock}{\relax}
\providecommand{\bibinfo}[2]{#2}
\providecommand{\BIBentrySTDinterwordspacing}{\spaceskip=0pt\relax}
\providecommand{\BIBentryALTinterwordstretchfactor}{4}
\providecommand{\BIBentryALTinterwordspacing}{\spaceskip=\fontdimen2\font plus
\BIBentryALTinterwordstretchfactor\fontdimen3\font minus
  \fontdimen4\font\relax}
\providecommand{\BIBforeignlanguage}[2]{{%
\expandafter\ifx\csname l@#1\endcsname\relax
\typeout{** WARNING: IEEEtran.bst: No hyphenation pattern has been}%
\typeout{** loaded for the language `#1'. Using the pattern for}%
\typeout{** the default language instead.}%
\else
\language=\csname l@#1\endcsname
\fi
#2}}
\providecommand{\BIBdecl}{\relax}
\BIBdecl

\bibitem{Nguyen2022}
D.~C. Nguyen, M.~Ding, P.~N. Pathirana, A.~Seneviratne, J.~Li, D.~Niyato,
  O.~Dobre, and H.~V. Poor, ``6{G} internet of things: A comprehensive
  survey,'' \emph{IEEE Internet Things J.}, vol.~9, no.~1, pp. 359--383, Jan.
  2022.

\bibitem{Z_Zhang2019}
Z.~Zhang, Y.~Xiao, Z.~Ma, M.~Xiao, Z.~Ding, X.~Lei, G.~K. Karagiannidis, and
  P.~Fan, ``6{G} wireless networks: vision, requirements, architecture, and key
  technologies,'' \emph{IEEE Veh. Technol. Mag.}, vol.~14, no.~3, pp. 28--41,
  Sep. 2019.

\bibitem{lingsheng_10925171}
L.~Meng, Y.~L. Guan, Y.~Ge, and Z.~Liu, ``Flag sequence set design for
  low-complexity delay-doppler estimation,'' \emph{IEEE Transactions on
  Vehicular Technology}, pp. 1--17, 2025.

\bibitem{liu}
F.~Liu, C.~Masouros, A.~P. Petropulu, H.~Griffiths, and L.~Hanzo, ``Joint radar
  and communication design: Applications, state-of-the-art, and the road
  ahead,'' \emph{IEEE Trans. Commun.}, vol.~68, no.~6, pp. 3834--3862, Jun.
  2020.

\bibitem{10012421}
Z.~Wei, H.~Qu, Y.~Wang, X.~Yuan, H.~Wu, Y.~Du, K.~Han, N.~Zhang, and Z.~Feng,
  ``Integrated sensing and communication signals toward 5{G}-{A} and 6{G}: A
  survey,'' \emph{IEEE Internet Things J.}, vol.~10, no.~13, pp.
  11\,068--11\,092, July 2023.

\bibitem{1638663}
T.~Wang, J.~Proakis, E.~Masry, and J.~Zeidler, ``Performance degradation of
  {OFDM} systems due to {D}oppler spreading,'' \emph{IEEE Trans. Wireless
  Commun.}, vol.~5, no.~6, pp. 1422--1432, Jun. 2006.

\bibitem{ICI_weighted}
Y.~Zhao and S.~G. Haggman, ``Intercarrier interference self-cancellation scheme
  for {OFDM} mobile communication systems,'' \emph{IEEE Trans. Commun.},
  vol.~49, no.~7, pp. 1185--1191, Jul. 2001.

\bibitem{R_Hadani}
R.~Hadani, S.~Rakib, M.~Tsatsanis, A.~Monk, A.~J. Goldsmith, A.~F. Molisch, and
  R.~Calderbank, ``Orthogonal time frequency space modulation,'' in \emph{Proc.
  IEEE Wireless Commun. Net. Conf. (WCNC)}, Mar. 2017, pp. 1--6.

\bibitem{9508932}
Z.~Wei, W.~Yuan, S.~Li, J.~Yuan, G.~Bharatula, R.~Hadani, and L.~Hanzo,
  ``Orthogonal time-frequency space modulation: A promising next-generation
  waveform,'' \emph{IEEE Wireless Commun.}, vol.~28, no.~4, pp. 136--144, Aug.
  2021.

\bibitem{10891132}
Q.~Deng, Y.~Ge, and Z.~Ding, ``A unifying view of {OTFS} and its many
  variants,'' \emph{IEEE Commun. Surv. Tuts.}, pp. 1--1, 2025.

\bibitem{9738478}
Z.~Wei, W.~Yuan, S.~Li, J.~Yuan, and D.~W.~K. Ng, ``Off-grid channel estimation
  with sparse bayesian learning for {OTFS} systems,'' \emph{IEEE Trans. on
  Wireless Commun.}, vol.~21, no.~9, pp. 7407--7426, Sep. 2022.

\bibitem{10791452}
J.~Wu, W.~Yuan, Z.~Wei, K.~Zhang, F.~Liu, and D.~Wing Kwan~Ng, ``Low-complexity
  minimum {BER} precoder design for {ISAC} systems: A delay-doppler
  perspective,'' \emph{IEEE Trans. on Wireless Commun.}, vol.~24, no.~2, pp.
  1526--1540, Feb. 2025.

\bibitem{OCDM}
X.~Ouyang and J.~Zhao, ``Orthogonal chirp division multiplexing,'' \emph{IEEE
  Trans. Commun.}, vol.~64, no.~9, pp. 3946--3957, Sep. 2016.

\bibitem{OCDM_performance}
M.~S. Omar and X.~Ma, ``Performance analysis of {OCDM} for wireless
  communications,'' \emph{IEEE Trans. Wireless Commun.}, vol.~20, no.~7, pp.
  4032--4043, Jul. 2021.

\bibitem{AFDM_TWC}
A.~Bemani, N.~Ksairi, and M.~Kountouris, ``Affine frequency division
  multiplexing for next generation wireless communications,'' \emph{IEEE Trans.
  Wireless Commun.}, vol.~22, no.~11, pp. 8214--8229, Nov. 2023.

\bibitem{luoqu}
Q.~Luo, P.~Xiao, Z.~Liu, Z.~Wan, N.~Thomos, Z.~Gao, and Z.~He, ``{AFDM-SCMA}: A
  promising waveform for massive connectivity over high mobility channels,''
  \emph{IEEE Trans. Wireless Commun.}, vol.~23, no.~10, pp. 14\,421--14\,436,
  Oct. 2024.

\bibitem{10711268}
K.~Zheng, M.~Wen, T.~Mao, L.~Xiao, and Z.~Wang, ``Channel estimation for {AFDM}
  with superimposed pilots,'' \emph{IEEE Trans. Veh. Technol.}, pp. 1--6, 2024.

\bibitem{yiwei}
Y.~Tao, M.~Wen, Y.~Ge, J.~Li, E.~Basar, and N.~Al-Dhahir, ``Affine frequency
  division multiplexing with index modulation: Full diversity condition,
  performance analysis, and low-complexity detection,'' \emph{IEEE J. Sel.
  Areas Commun.}, vol.~43, no.~4, pp. 1041--1055, Apr. 2025.

\bibitem{W_Benzine_CS}
W.~Benzine, A.~Bemani, N.~Ksairi, and D.~Slock, ``Affine frequency division
  multiplexing for compressed sensing of time-varying channels,'' in
  \emph{Proc. IEEE 25th Int. Workshop Signal Process. Adv. Wireless Commun.
  (SPAWC)}, Sep. 2024.

\bibitem{AFDM_MIMO}
H.~Yin, X.~Wei, Y.~Tang, and K.~Yang, ``Diagonally reconstructed channel
  estimation for {MIMO-AFDM} with inter-doppler interference in doubly
  selective channels,'' \emph{IEEE Trans. Wireless Commun.}, vol.~23, no.~10,
  pp. 14\,066--14\,079, Oct. 2024.

\bibitem{K_R_R_Ranasinghe}
K.~R.~R. Ranasinghe, H.~S. Rou, G.~T.~F. De~Abreu, T.~Takahashi, and K.~Ito,
  ``Joint channel, data and radar parameter estimation for {AFDM} systems in
  doubly-dispersive channels,'' \emph{IEEE Trans. Wireless Commun.}, 2024,
  early access.

\bibitem{bistatic_AFDM}
J.~Zhu, Y.~Tang, F.~Liu, X.~Zhang, H.~Yin, and Y.~Zhou, ``{AFDM}-based bistatic
  integrated sensing and communication in static scatterer environments,''
  \emph{IEEE Commun. Lett.}, vol.~13, no.~8, pp. 2245--2249, Aug. 2024.

\bibitem{Y_Ni_ISAC}
Y.~Ni, Z.~Wang, P.~Yuan, and Q.~Huang, ``An {AFDM}-based integrated sensing and
  communications,'' in \emph{Proc. Int. Sym. Wireless Commun. Syst. (ISWCS)},
  Oct 2022, pp. 1--6.

\bibitem{ML}
A.~Bemani, N.~Ksairi, and M.~Kountouris, ``Integrated sensing and
  communications with affine frequency division multiplexing,'' \emph{IEEE
  Wireless Commun. Lett.}, vol.~13, no.~5, pp. 1255--1259, May 2024.

\bibitem{AFDM_ICC}
Y.~Luo, Y.~Guan, Y.~Ge, and C.~Yuen, ``Target sensing with off-grid sparse
  bayesian learning for {AFDM-ISAC} system,'' \emph{arXiv:2503.10011}, 2025.

\bibitem{Z_Xiao}
Z.~Xiao, R.~Liu, M.~Li, Q.~Liu, and A.~L. Swindlehurst, ``A novel joint
  angle-range-velocity estimation method for {MIMO-OFDM} {ISAC} systems,''
  \emph{IEEE Trans. Signal Process.}, vol.~72, pp. 3805--3818, Aug. 2024.

\bibitem{two_stage}
L.~Leyva, D.~Castanheira, A.~Silva, and A.~Gameiro, ``Two-stage estimation
  algorithm based on interleaved {OFDM} for a cooperative bistatic {ISAC}
  scenario,'' in \emph{Proc. IEEE 95th Veh. Technol. Conf. (VTC-Spring)}, Jun.
  2022, pp. 1--6.

\bibitem{ESPRIT}
Y.~Xiang, Y.~Gao, X.~Yang, S.~Kang, and M.~Shao, ``An {ESPRIT}-based moving
  target sensing method for {MIMO-OFDM} {ISAC} system,'' \emph{IEEE Commun.
  Lett.}, vol.~27, no.~12, pp. 3205--3209, Dec. 2023.

\bibitem{Z_Gao}
Z.~Gao, C.~Hu, L.~Dai, and Z.~Wang, ``Channel estimation for millimeter-wave
  massive {MIMO} with hybrid precoding over frequency-selective fading
  channels,'' \emph{IEEE Commun. Lett.}, vol.~20, no.~6, pp. 1259--1262, Jun.
  2016.

\bibitem{J_Lee}
J.~Lee, G.-T. Gil, and Y.~H. Lee, ``Channel estimation via orthogonal matching
  pursuit for hybrid {MIMO }systems in millimeter wave communications,''
  \emph{IEEE Trans. Commun.}, vol.~64, no.~6, pp. 2370--2386, Jun. 2016.

\bibitem{low_rank}
Z.~Zhou, J.~Fang, L.~Yang, H.~Li, Z.~Chen, and R.~S. Blum, ``Low-rank tensor
  decomposition-aided channel estimation for millimeter wave {MIMO-OFDM}
  systems,'' \emph{IEEE J. Sel. Areas Commun.}, vol.~35, no.~7, pp. 1524--1538,
  Jul. 2017.

\bibitem{Y_Lin}
Y.~Lin, S.~Jin, M.~Matthaiou, and X.~You, ``Tensor-based channel estimation for
  millimeter wave {MIMO-OFDM} with dual-wideband effects,'' \emph{IEEE Trans.
  Commun.}, vol.~68, no.~7, pp. 4218--4232, Jul. 2020.

\bibitem{ruoyu}
R.~Zhang, L.~Cheng, S.~Wang, Y.~Lou, W.~Wu, and D.~W.~K. Ng, ``Tensor
  decomposition-based channel estimation for hybrid mmwave massive {MIMO} in
  high-mobility scenarios,'' \emph{IEEE Trans. Commun.}, vol.~70, no.~9, pp.
  6325--6340, Sep. 2022.

\bibitem{D_Nion}
D.~Nion and N.~D. Sidiropoulos, ``Tensor algebra and multidimensional harmonic
  retrieval in signal processing for {MIMO} radar,'' \emph{IEEE Trans. Signal
  Process.}, vol.~58, no.~11, pp. 5693--5705, Nov. 2010.

\bibitem{M_Cao}
M.~Cao, S.~A. Vorobyov, and A.~Hassanien, ``Transmit array interpolation for
  {DOA} estimation via tensor decomposition in {2-D MIMO} radar,'' \emph{IEEE
  Trans. Signal Process.}, vol.~65, no.~19, pp. 5225--5239, Oct. 2017.

\bibitem{tensor_OTFS}
X.~Li, B.~Chang, and Z.~Chen, ``Tensor decomposition based {THz} channel
  estimation in {OTFS} for integrated sensing and communications,'' in
  \emph{Proc. IEEE Global Comm. Conf.}, Dec. 2023, pp. 3996--4001.

\bibitem{ruoyu_ISAC}
R.~Zhang \emph{et~al.}, ``Integrated sensing and communication with massive
  {MIMO}: A unified tensor approach for channel and target parameter
  estimation,'' \emph{IEEE Trans. Wireless Commun.}, vol.~23, no.~8, pp.
  8571--8587, Aug. 2024.

\bibitem{yirui}
Y.~Luo, Y.~Guan, and E.~Gunawan, ``Uplink sensing with unknown transmitter
  position in clutter environment via tensor decomposition,'' in \emph{Proc.
  IEEE 97th Veh. Technol. Conf. (VTC-Spring)}, Jun. 2023, pp. 1--5.

\bibitem{near_dai}
M.~Cui and L.~Dai, ``Channel estimation for extremely large-scale {MIMO}:
  Far-field or near-field?'' \emph{IEEE Trans. Commun.}, vol.~70, no.~4, pp.
  2663--2677, Jan. 2022.

\bibitem{L_Wei}
L.~Wei, C.~Huang, G.~C. Alexandropoulos, Z.~Yang, J.~Yang, W.~E.~I. Sha,
  Z.~Zhang, M.~Debbah, and C.~Yuen, ``Tri-polarized holographic {MIMO} surfaces
  for near-field communications: Channel modeling and precoding design,''
  \emph{IEEE Trans. Wireless Commun.}, vol.~22, no.~12, pp. 8828--8842, Dec.
  2023.

\bibitem{10149471}
Y.~Pan, C.~Pan, S.~Jin, and J.~Wang, ``{RIS}-aided near-field localization and
  channel estimation for the terahertz system,'' \emph{IEEE J. Sel. Topics
  Signal Process.}, vol.~17, no.~4, pp. 878--892, Jul. 2023.

\bibitem{10135096}
Z.~Wang, X.~Mu, and Y.~Liu, ``Near-field integrated sensing and
  communications,'' \emph{IEEE Commun. Lett.}, vol.~27, no.~8, pp. 2048--2052,
  Aug. 2023.

\bibitem{10579914}
H.~Li, Z.~Wang, X.~Mu, P.~Zhiwen, and Y.~Liu, ``Near-field integrated sensing,
  positioning, and communication: A downlink and uplink framework,'' \emph{IEEE
  J. Sel. Areas Commun.}, vol.~42, no.~9, pp. 2196--2212, Sep. 2024.

\bibitem{W_Zuo}
W.~Zuo, J.~Xin, N.~Zheng, and A.~Sano, ``Subspace-based localization of
  far-field and near-field signals without eigendecomposition,'' \emph{IEEE
  Trans. Signal Process.}, vol.~66, no.~17, pp. 4461--4476, Sep. 2018.

\bibitem{8753714}
Z.~Zheng, M.~Fu, W.-Q. Wang, S.~Zhang, and Y.~Liao, ``Localization of mixed
  near-field and far-field sources using symmetric double-nested arrays,''
  \emph{IEEE Trans. on Antennas and Propagation}, vol.~67, no.~11, pp.
  7059--7070, Nov. 2019.

\bibitem{Kruskal}
J.~B. Kruskal, ``Three-way arrays: Rank and uniqueness or trilinear
  decompositions, with applications to arithmetic complexity and statistics,''
  \emph{Linear Algebra Appl.}, vol.~18, no.~2, pp. 95--138, 1977.

\bibitem{van_cpd}
M.~Sørensen and L.~D. Lathauwer, ``Blind signal separation via tensor
  decomposition with vandermonde factor: Canonical polyadic decomposition,''
  \emph{IEEE Trans. Signal Process.}, vol.~61, no.~22, pp. 5507--5519, Aug.
  2013.

\bibitem{995060}
E.~Fishler, M.~Grosmann, and H.~Messer, ``Detection of signals by information
  theoretic criteria: General asymptotic performance analysis,'' \emph{IEEE
  Trans. Signal Process.}, vol.~50, no.~5, pp. 1027--1036, May 2002.

\bibitem{tensor}
T.~Kolda and B.~Bader, ``Tensor decompositions and applications,'' \emph{SIAM
  Rev.}, vol.~51, no.~3, pp. 455--500, Sep. 2009.

\bibitem{CRLB}
S.~M. Kay, \emph{Fundamentals of Statistical Signal Processing: Estimation
  Theory}.\hskip 1em plus 0.5em minus 0.4em\relax Prentice Hall, 1993.

\bibitem{CRLB_cpd}
X.~Liu and N.~D. Sidiropoulos, ``Cramer-{R}ao lower bounds for low-rank
  decomposition of multidimensional arrays,'' \emph{IEEE Trans. Signal
  Process.}, vol.~49, no.~9, pp. 2074--2086, Sep. 2001.

\end{thebibliography}
\end{document}